\documentclass[article,longauth]{aa} 
\usepackage{graphicx}
\usepackage{txfonts}
\usepackage[colorlinks=true,linkcolor=blue,citecolor=blue]{hyperref} 

\usepackage{xcolor}
\usepackage{float}
\usepackage{soul}
\usepackage{orcidlink}
\usepackage{natbib}
\bibpunct{(}{)}{;}{a}{}{,}

\begin{document} 

\title{First unambiguous detection of ammonia in the atmosphere of a planetary mass companion with JWST/MIRI coronagraphs}

\subtitle{}

\author{
    Mathilde Mâlin\orcidlink{0000-0002-2918-8479}\inst{\ref{jhu},\ref{stsci},\ref{lesia}},
    Anthony Boccaletti\orcidlink{0000-0001-9353-2724}\inst{\ref{lesia}}, 
    Clément Perrot\orcidlink{0000-0003-3831-0381}\inst{\ref{lesia}},
    Pierre Baudoz\orcidlink{0000-0002-2711-7116}\inst{\ref{lesia}}, 
    Daniel Rouan\orcidlink{0000-0002-2352-1736}\inst{\ref{lesia}},
    Pierre-Olivier Lagage\inst{\ref{cea}},
    Rens Waters\orcidlink{0000-0002-5462-9387}\inst{\ref{radboud},\ref{hfml},\ref{sron}},
    Manuel G\"udel\orcidlink{0000-0001-9818-0588}\inst{\ref{vienna},\ref{eth}},
    Thomas Henning\orcidlink{0000-0002-1493-300X}\inst{\ref{mpia}},
    Bart Vandenbussche\orcidlink{0000-0002-1368-3109}\inst{\ref{leuven}},
    Olivier Absil\orcidlink{0000-0002-4006-6237}\inst{\ref{star}},
    David Barrado\orcidlink{0000-0002-5971-9242}\inst{\ref{cab}},
    Benjamin Charnay\inst{\ref{lesia}},
    Elodie Choquet\orcidlink{0000-0002-9173-0740}\inst{\ref{lam}},
    Christophe Cossou\orcidlink{0000-0001-5350-4796}\inst{\ref{parissaclay}}, 
    Camilla Danielski\orcidlink{0000-0002-3729-2663}\inst{\ref{inaf}},
    Leen Decin\orcidlink{0000-0002-5342-8612}\inst{\ref{leuven}},
    Adrian M. Glauser\orcidlink{0000-0001-9250-1547}\inst{\ref{eth}},
    John Pye\orcidlink{0000-0002-0932-4330}\inst{\ref{leicester}},
    Goran Olofsson\orcidlink{0000-0003-3747-7120}\inst{\ref{stockholm}},
    Alistair Glasse\orcidlink{0000-0002-2041-2462}\inst{\ref{ukatc}},
    Polychronis Patapis\orcidlink{0000-0001-8718-3732}\inst{\ref{eth}},
    Pierre Royer\orcidlink{0000-0001-9341-2546}\inst{\ref{leuven}},
    Silvia Scheithauer\orcidlink{0000-0003-4559-0721}\inst{\ref{mpia}},
    Eugene Serabyn\inst{\ref{jpl}},
    Pascal Tremblin\inst{\ref{cea},\ref{parissaclay2}},
    Niall Whiteford\orcidlink{0000-0001-8818-1544}\inst{\ref{museum}},
    Ewine F. van Dishoeck\orcidlink{0000-0001-7591-1907}\inst{\ref{leiden}},
    G\"oran Ostlin\orcidlink{0000-0002-3005-1349}\inst{\ref{oskar}},
    Tom P.\ Ray\orcidlink{0000-0002-2110-1068}\inst{\ref{dublin}}
    Gillian Wright\orcidlink{0000-0001-7416-7936}\inst{\ref{ukatc}}
    }
\institute{Department of Physics \& Astronomy, Johns Hopkins University, 3400 N. Charles Street, Baltimore, MD 21218, USA\label{jhu}
\and Space Telescope Science Institute, 3700 San Martin Drive, Baltimore, MD 21218, USA\label{stsci}
\and LIRA, Observatoire de Paris, Universit{\'e} PSL, Sorbonne Universit{\'e}, Universit{\'e} Paris Cité, CY Cergy Paris Universit{\'e}, CNRS, 92190 Meudon, France\label{lesia} 
\and Universit{\'e} Paris-Saclay, Universit{\'e} Paris Cit{\'e}, CEA, CNRS, AIM, 91191, Gif-sur-Yvette, France\label{cea}
\and  Department of Astrophysics/IMAPP, Radboud University, PO Box 9010, 6500 GL Nijmegen, the Netherlands\label{radboud}
\and  HFML - FELIX. Radboud University PO box 9010, 6500 GL Nijmegen, the Netherlands\label{hfml}
\and  SRON Netherlands Institute for Space Research, Niels Bohrweg 4, 2333 CA Leiden, the Netherlands\label{sron}
\and  Department of Astrophysics, University of Vienna, T\"urkenschanzstrasse 17, 1180 Vienna, Austria\label{vienna}
\and ETH Zürich, Institute for Particle Physics and Astrophysics, Wolfgang-Pauli-Strasse 27, 8093 Zürich, Switzerland\label{eth}
\and  Max-Planck-Institut f\"ur Astronomie (MPIA), K\"onigstuhl 17, 69117 Heidelberg, Germany \label{mpia}
\and Institute of Astronomy, KU Leuven, Celestijnenlaan 200D, 3001 Leuven, Belgium\label{leuven}
\and STAR Institute, Universit\'e de Li\`ege, All\'ee du Six Ao\^ut 19c, 4000 Li\`ege, Belgium\label{star}
\and Centro de Astrobiología (CAB), CSIC-INTA, ESAC Campus, Camino Bajo del Castillo s/n, 28692 Villanueva de la Cañada, Madrid, Spain \label{cab}
\and Aix Marseille Univ, CNRS, CNES, LAM, Marseille, France\label{lam}
\and Université Paris-Saclay, CEA, IRFU, 91191, Gif-sur-Yvette, France\label{parissaclay}
\and INAF - Osservatorio Astrofisico di Arcetri, Largo E. Fermi 5, 50125, Firenze, Italy\label{inaf}
\and  School of Physics \& Astronomy, 
Space Park Leicester, University of Leicester, 92 Corporation Road, Leicester, LE4 5SP, UK\label{leicester}
\and Department of Astronomy, Stockholm University, AlbaNova University Center, 10691 Stockholm, Sweden\label{stockholm}
\and UK Astronomy Technology Centre, Royal Observatory, Blackford Hill, Edinburgh EH9 3HJ, UK\label{ukatc}
\and Jet Propulsion Laboratory, California Institute of Technology, 4800 Oak Grove Dr.,Pasadena, CA 91109, USA\label{jpl}
\and Université Paris-Saclay, UVSQ, CNRS, CEA, Maison de la Simulation, 91191, Gif-sur-Yvette, France\label{parissaclay2}
\and Department of Astrophysics, American Museum of Natural History, New York, NY 10024, USA\label{museum}
\and Leiden Observatory, Leiden University, P.O. Box 9513, 2300 RA Leiden, the Netherlands\label{leiden}
\and Department of Astronomy, Oskar Klein Centre, Stockholm University, 106 91 Stockholm, Sweden\label{oskar}
\and School of Cosmic Physics, Dublin Institute for Advanced Studies, 31 Fitzwilliam Place, Dublin, D02 XF86, Ireland\label{dublin} }

\date{}
    \abstract
   {The newly accessible mid-infrared (MIR) window offered by the \textit{James Webb} Space Telescope (JWST) for exoplanet imaging is expected to provide valuable information to characterize their atmospheres.
   In particular, coronagraphs on board the JWST Mid-InfraRed instrument (MIRI) are capable of imaging the coldest directly imaged giant planets at the wavelengths where they emit most of their flux.
   The MIRI coronagraphs have been specially designed to detect the NH$_3$ absorption around 10.5 $\mu$m, which has been predicted by atmospheric models and should be detectable for planets colder than 1200\,K.}
   {We aim to assess the presence of NH$_3$ while refining the atmospheric parameters of one of the coldest companions detected by directly imaging GJ\,504\,b.
   Its mass is still a matter of debate and depending on the host star age estimate, the companion could either be placed in the brown dwarf regime of $\sim$ 20 M$_{\rm Jup}$ or in the young Jovian planet regime of $\sim$ 4 M$_{\rm Jup}$.}
   {We present an analysis of new MIRI observations, using the coronagraphic filters F1065C, F1140C, and F1550C of the GJ\,504 system.
   We took advantage of previous observations of reference stars to build a library of images and to perform a more efficient subtraction of the stellar diffraction pattern.
   We used an atmospheric grid from the \texttt{Exo-REM} model to refine the atmospheric parameters by combining archival near-infrared (NIR) photometry with the MIR photometry.}
   {We detected the presence of NH$_3$ at 
   12.5\,$\sigma$ and measured its volume mixing ratio of 10$^{-5.3 \pm 0.07}$ 
   in the atmosphere of GJ\,504\,b. These results are in line with atmospheric model expectations for a planetary-mass object and observed in brown dwarfs within a similar temperature range.
   The best-fit model with \texttt{Exo-REM} provides updated values of its atmospheric parameters, yielding a temperature of T$_\text{eff}$ = 512$\pm$10\,K
   and radius of R = 1.08$^{+0.04}_{-0.03}$ R$_{\rm Jup}$.}
   {These observations demonstrate the capability of MIRI coronagraphs to detect NH$_3$ and to provide the first MIR observations of one of the coldest directly imaged companions. Overall,
   NH$_3$ is a key molecule for characterizing the  atmospheres of cold planets, offering valuable insights into their surface gravity.
   These observations provide valuable information for future spectroscopic observations planned with JWST, in particular, with the MIRI medium-resolution spectrometer (MRS), which will allow us to characterize the atmosphere of GJ\,504\,b in depth.}
\authorrunning{M. Mâlin et al.}
\titlerunning{First unambiguous detection of ammonia in the atmosphere of a planetary mass companion.}
\maketitle

\section{Introduction}
\begin{figure}[h]
    \centering
    \includegraphics[width=8.5cm]{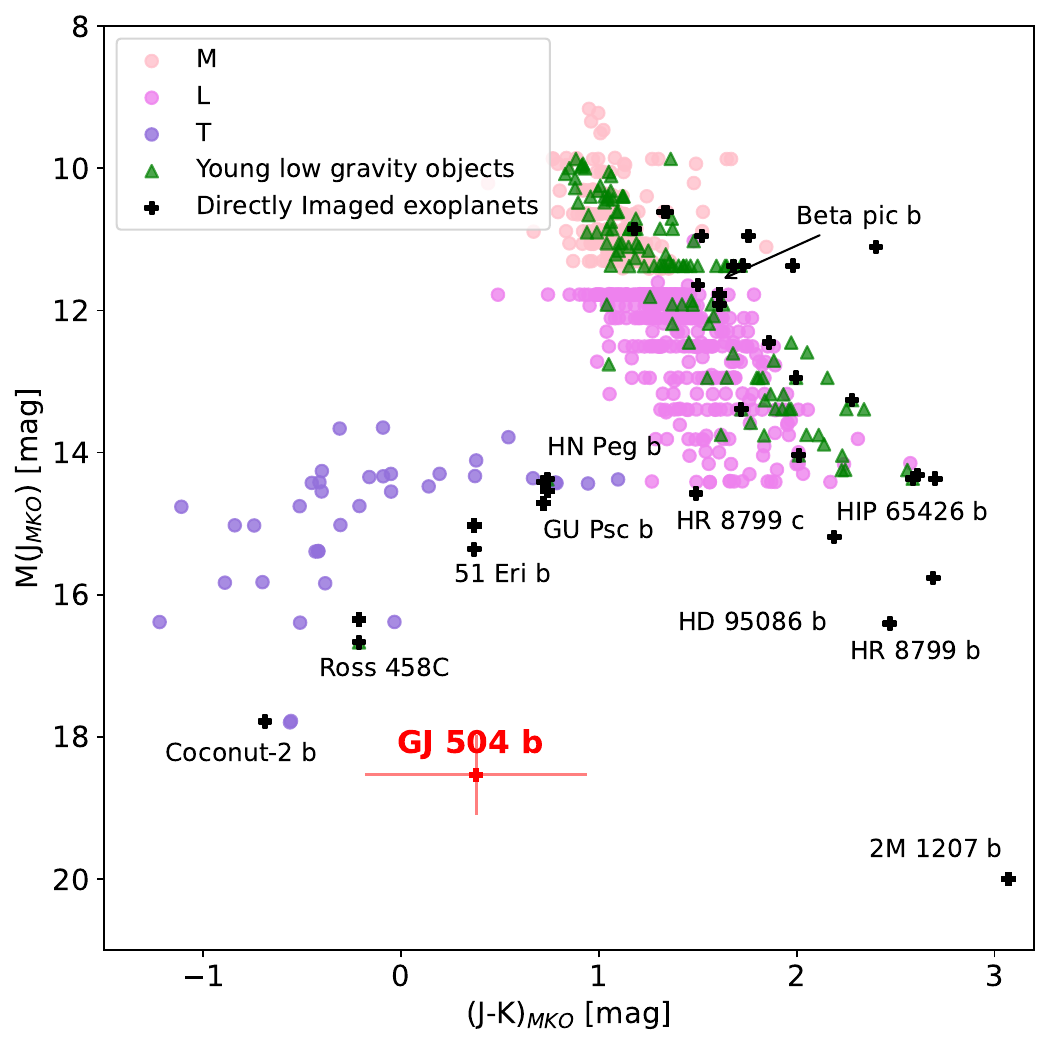}
    \caption{Color-magnitude diagram of the field brown dwarfs (M, L, and T dwarfs are shown in color), objects with confirmed youth or low gravity, and well-known directly imaged exoplanets \citep[photometry from the UltracoolSheet,][]{best_ultracoolsheet_2024}.
    GJ\,504\,b is at a unique position between the sequence of young low-gravity objects and old T-type brown dwarfs.}
    \label{fig:CMD_exoplanet_GJ504}
\end{figure}

The planetary-mass companion GJ\,504\,b is one of the few imaged planets to bridge the gap between the population of directly imaged young warm exoplanets $\sim$\,1000\,K and our Solar System's Jupiter at $\sim$\,130\,K.
Together with recent imaging of the planet Eps\,Ind\,b at $\sim$\,275\,K \citep{matthews_temperate_2024}, GJ\,504\,b is one of the coldest planetary-mass companions ($\sim$\,500\,K) imaged to date.
It orbits a solar-type star \citep[spectral type G0V,][]{anderson_detectability_2010} at a separation of 43 au, showcasing an orbit that is slightly beyond that of the nearby Neptune (i.e., 30\,au).
GJ\,504\,b exhibits bluer colors in the near-infrared (NIR) than any previously directly imaged exoplanet \citep[$J - H = -0.23\,\text{mag}$,][]{kuzuhara_direct_2013}, but is redder than any observed brown dwarfs with a similar temperature or brightness, probing an unexplored parameter space of the color-magnitude diagram \citep{bonnefoy_gj_2018}. 
This is shown in Fig. \ref{fig:CMD_exoplanet_GJ504} with the color-magnitude diagram at NIR wavelengths.
Its properties are rather typical of late T-type brown dwarfs, suggesting a largely cloud-free atmosphere.
Indeed, \cite{janson_direct_2013} confirmed the first detection of CH$_4$ in its atmosphere, as expected for a T-type object, which is hints at the presence of disequilibrium chemistry in its atmosphere.

Furthermore, the existence of this object represents a challenge for theories of planetary formation and evolution.
To date, neither  of the two most well-established planet formation scenarios can explain GJ\,504\,b's current orbital separation and super-solar metallicity, raising fundamental questions about its origin \citep{skemer_leech_2016, bonnefoy_gj_2018}.
In the core accretion model, planets form close to their stars, with a predicted typical outer boundary of $\sim$ 30 au. 
Forming planets slowly accrete dust, grains, and pebbles over a few millions of years, after which they accrete their gaseous envelope \citep{bodenheimer_models_2000}. 
This scenario fails to explain the large orbital separation of GJ\,504\,b.
The disk instability model mirrors the process of star and brown dwarf formation: planets are believed to form through the collapse of a massive protoplanetary disk due to gravitational instabilities, leading to fragmentation and subsequent planet formation \citep{marley_luminosity_2007}.
According to this model, planets are presumed to retain the chemical composition of the disk material at the location of their formation, meaning that the planet is expected to have the same metallicity as the host-star \citep{oberg_effects_2011,molliere_interpreting_2022}.
The super-solar metallicity observed in GJ\,504\,b, along with its derived low mass of 4\,M$_{\rm Jup}$ \citep{kuzuhara_direct_2013}
cannot be explained with the disk instability model, unless it undergoes significant migration either inward or outward \citep{shibata_origin_2020, turrini_tracing_2021}.
The metallicity of the companion is more enriched ([M/H] $\approx$ +0.6) in metal than its parent star GJ\,504\,A \citep[{[M/H]} $\approx$ +0.1 to 0.3,][]{skemer_leech_2016}.
Given that its atmosphere and the history of its formation and evolution are not well understood, previous studies have given rise to controversy over its planetary nature.
Indeed, its derived mass relies on evolutionary models, based on the assumed age for this system.

The stellar gyrochronological and chromospheric activities of GJ\,504\,A have led to the determination of its characteristics as a rather young star \citep[160$^{+350}_{-60}$ Myr,][]{kuzuhara_direct_2013}.
Later studies indicated an older stellar age, between 1.5 and 4\,Gyr: \cite{fuhrmann_age_2015} and \cite{dorazi_critical_2017} suggested that the high levels of rotation and chromospheric activity, normally characteristic of a young stellar age, are due to the recent engulfment of a short-period hot Jupiter, further arguing in favor of an older system.
Isochronal studies also show two solutions corresponding to the star: one being at a young age and the other at an older age \citep{bonnefoy_gj_2018}.
Finally, more recent studies favor the hypothesis of a young system, based on high-resolution spectra of the star and indicators of stellar activity \citep{subjak_search_2023, di_mauro_characterization_2022}. 
However, they did not rule out the hypothesis of an older system.
The degeneracies in the system's age have led to highly disparate mass values for GJ\,504\,b; namely:  $M = 1.3 ^{+0.6}_{-0.3}\,M_{\rm Jup}$ or $ M = 23.3^{+10}_{-9}\,M_{\rm Jup}$
for the ages of young and old isochronal systems of 21$\pm$2\,Myr and 4.0$\pm$1.8\,Gyr, respectively \citep{bonnefoy_gj_2018}.
%
These values allow for such an object to be either one of the lowest-mass young planetary imaged companions known to date or, simply, an old brown dwarf.

In this work, we present the analysis of new observations from the \textit{James Webb} Space Telescope (JWST) Mid-InfraRed instrument (MIRI)  obtained within the framework of the ExoMIRI Guaranteed Time Observations (GTO) program 1277 (PI: P.-O. Lagage).
They are also part of MIRIco, a EU/US coordinated observing effort with the MIRI coronagraphs between programs 1194, 1277, and 1241.
In Sect. \ref{sec:obs_data_red}, we present the observational parameters, data reduction, and use of a reference star library.
In Sect. \ref{sec:atm_charact}, we describe  the atmospheric characterization of the object.
We discuss our results in Sect. \ref{sec:discussion} and our conclusions in Sect. \ref{sec:conclusion}.

\section{Observations and data reduction}
\label{sec:obs_data_red}
\subsection{Program observations}
The system was observed with all three of MIRI's 4-Quadrant Phase-Masks coronagraphs \citep[4QPM,][]{rouan_fourquadrant_2000}, along with the paired filters F1065C, F1140C, and F1550C.
These filters are centered at 10.575, 11.30, and 15.50\,$\mu$m, with a $\sim$ 5\% bandwidth, proving the first mid-infrared (MIR) images of the system.
Background observations are included for each filter to mitigate the “glowstick” effect identified during commissioning \citep{boccaletti_jwstmiri_2022}. 
The background is observed using two dithers, which are then averaged to optimize its subtraction.
No dedicated reference stars were observed during this sequence of observations.
In fact, the purpose of this GTO sequence was precisely to test to which level of contrast the diffraction pattern can be subtracted out with one or several other reference stars from our program or from other programs. With a favorable separation of $\sim$\,2.5$''$, we were expecting the companion to be detected even if starlight subtraction would be degraded.
As a result, we anticipate that stellar subtraction may not be optimal; however, this does not hinder the ability to detect the companion.
The observation parameters are summarized in Table \ref{tab:log_obs}. 
The system is observed in the F1065C filter with twice more integrations (720\,s instead of 360\,s for the F1140 and F1550C filters), as GJ\,504\,b is expected to appear fainter at this specific wavelength, due to ammonia absorption.
\begin{table*}[h!]
\caption{Parameters of the observations of the GJ\,504 system.}
\begin{center}
\begin{tabular}{lllllllll}
\hline
\hline
Date and time  UT   & Filter & Object   &  Type     & Obs ID  & $N_{\rm group}$ & $N_{\rm int}$ & N$_{\rm dither}$ & $T_{\rm exp}$ (s)                          \\
        &       &        &          &          &             &           &        &   \\  
\hline
\hline
Jul 4, 2023 11:01:15 & F1065C & GJ\,504    & Target on    & obs 13  & 500 & 6  & 1 & 720.238 \\
Jul 4, 2023 11:45:50 & F1140C & GJ\,504    & Target on    & obs 14 & 500 & 3  & 1 & 359.999\\
Jul 4, 2023 12:17:52 & F1550C & GJ\,504    & Target on    & obs 15  & 500 & 3  & 1 & 359.999\\
\hline
Jul 4, 2023 12:41:07 & F1550C &   --  & background   & obs 16  & 500 & 2  & 2 & 479.839 \\
Jul 4, 2023 13:05:48 & F1140C &  --   & background    & obs 17 &500 & 2  & 2 & 479.839 \\
Jul 4, 2023 13:23:22 & F1065C &   --  & background    & obs 18  & 500 & 3  & 2 & 719.999 \\
\hline
\end{tabular}
\end{center}
\tablefoot{Date and time represent the starting time of the observation on the target, followed by the filter, the name of the object, the type, and the ID of each observation. 
The last parameters represent the observational parameters: number of groups, number of integration, number of dither positions, and the total exposure time.}
\label{tab:log_obs}
\end{table*}

\subsection{Data reduction}
\label{sec:data_red}
The uncalibrated data were retrieved (\texttt{$\_$uncal} files) from the Mikulski Archive for Space Telescopes MAST \citep{marston_overview_2018}.
%
The data reduction in this work was carried out in a similar way to that of previous datasets, as described in \cite{boccaletti_imaging_2024} and \cite{malin_unveiling_2024}.
We ran stage 1 of \texttt{JWST} pipeline\footnote{\href{https://jwst-pipeline.readthedocs.io/en/latest/jwst/package_index.html}{jwst-pipeline.readthedocs.io}, version: 1.12.5, CRDS = 1140} \citep{bushouse_jwst_20245}, which applies essential detector-level corrections to all exposure types to obtain a corrected count-rate image (\texttt{$\_$rates} files). 
Then, we ran stage 2 to subtract the background contribution and apply photometric calibration (\texttt{$\_$cal} files).
The flat-field correction is skipped to avoid increasing noise and the glowstick effect, as attenuation is more important at the edge of each quadrant. We checked that the impact on the photometry was lower than 2\%, so much smaller than the other sources of noise \citep{boccaletti_imaging_2024}.
Finally, we applied a $\sigma$-clipping function to correct the remaining bad pixels and NaN values.
Any pixel with a value greater than 3\,$\sigma$ compared to the median of its closest neighbors was replaced with this median value.
The images were rotated to align north with the top of the image using the position angle (angle of $\sim$ 121.7$^\circ$ for the F1065C filter and 118.5$^\circ$ for the F1140C and F1550C filters).
The MIRI coronagraphic images are fully dominated by diffraction, which we aim to subtract, as described in the following sections (see Sect. \ref{sec:stellar_sub}).

\begin{figure*}[h]
    \centering
    \includegraphics[width=18cm]{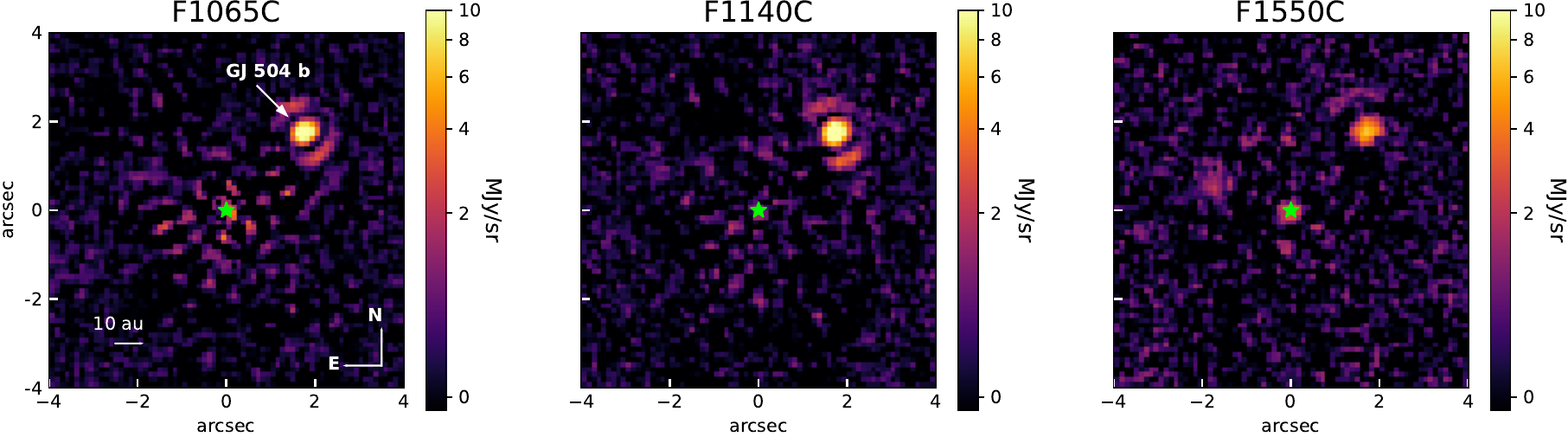}
    \caption{Coronagraphic images of GJ\,504 in each filter (F1065, F1140, and F1550) after the subtraction of an optimized-reference star.
    The coronagraphic center is illustrated with the small star in green and GJ\,504\,b is the bright source in the upper right corner, as indicated by the arrow in the F1065C image (first panel). 
    An asinh color scale is used to show more details. Images in linear scale are available in the appendix.} 
    \label{fig:image_sub}
\end{figure*}

\subsection{Building a library of references}
\label{sec:library}
We built a library of reference star observations to estimate and remove the stellar diffraction pattern.
We used all the available observations, at the time of this work (all reference star observations are publicly available in MAST), obtained with MIRI coronagraphic mode during the commissioning program 1037,  ERS program 1386, and the GTO and GO programs (i.e., GTO programs 1277, 1194, 1413, 1411, and 1241, along with  GO Cycle 1 programs 1668, 2153, 2243, and 2538). 
We also added reference observations from the GO cycle 2 available until June 1, 2024 (programs from 3254 and 3662).
This information is summarized in the appendix.
The result is a total of 10 stars observed in the F1065C filter, 18 stars in the F1140C, and 10 stars in the F1550C. 
We note that  filter F1140C is the most requested one in MIRI observation programs.
All reference star observations were reduced in the exact same way as described in Sect. \ref{sec:data_red}.
Each reference star has a spectral type that is similar to that of the target of each of the above observing programs, and it does not necessarily correspond to the spectral type of GJ\,504\,A.
All the reference stars observations were carried out using the small-grid-dither (SGD) strategy, which includes a small offset (by steps of 10 mas) between each of the dithered observations to account for the fact that the coronagraph center is not perfectly measured \citep{lajoie_small-grid_2016}. 
Most reference stars are observed with nine dithers, but some observational data sets use only five dithers.
Taking these different dither positions into account, there are a total of 70, 146, and 66 observations in the F1065C, F1140C, and F1550C filters, respectively.
\subsection{Stellar diffraction subtraction}
\label{sec:stellar_sub}
\begin{figure}[h]
    \centering
    \includegraphics[width=9cm]{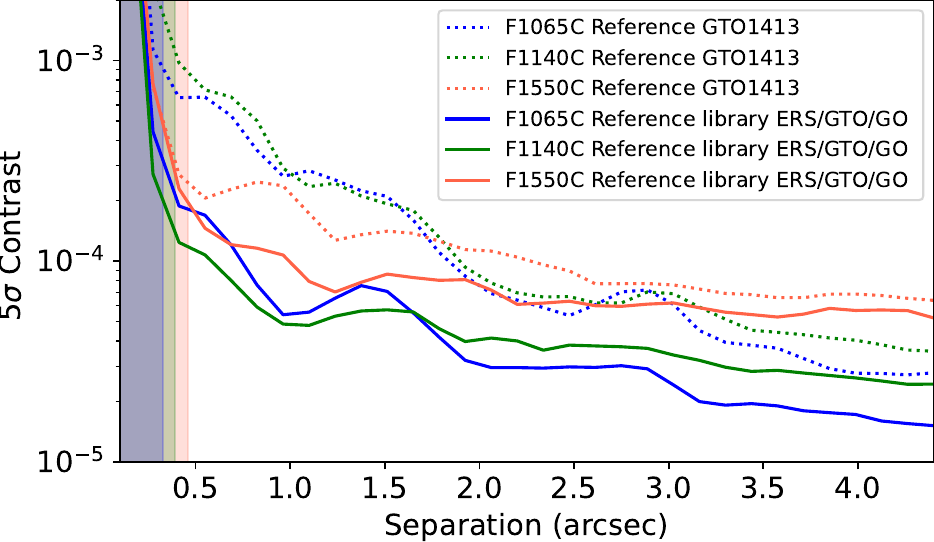}
    \caption{Contrast curve at 5$\sigma$ computed for each filter image. 
    All curves are computed using the 4Q-PCA stellar subtraction method.
    The plain lines correspond to the subtraction with the entire library and the dashed lines correspond to the case of using only the reference star from the GTO\,1413 program.
    The shaded regions correspond to the inner working angle of the coronagraphs.}
    \label{fig:contrast_curve_best}
\end{figure}

Using the reference star library (described in Sect. \ref{sec:library}), 
we used a principal component analysis (PCA) to reconstruct an optimized reference image.
We choose the number of PCA components to remove  to optimize the signal-to-noise ratio ($S/N$) for GJ\,504\,b.
%
We noticed that the stellar residuals can be quadrant-dependent, as illustrated in appendix 
, whereby  one quadrant has a larger amount of flux compared to the other three.
Therefore, we independently apply the PCA method to each of the four quadrants (method referred to as 4Q-PCA) to achieve the most effective subtraction.
This final result is shown in Fig. \ref{fig:image_sub} using the complete library of reference. 
The contrast limits curves for each filter are shown in Fig. \ref{fig:contrast_curve_best} (plain line).
They represent the achievable contrast sensitivity at 5\,$\sigma$ as a function of the separation for each filter. 
We also show the contrast limits obtained with a single reference image for comparison (Fig. \ref{fig:contrast_curve_best}, dashed line). 
Using a larger reference library provides a gain of at least one order of magnitude in terms of contrast limits at all separations with the F1065C and F1140C.
The gain in contrast is even larger at shorter separations.

Even if some stellar diffraction residuals persist at a shorter distance from the coronagraph center, 
the contrast sensitivity achieved at 5\,$\sigma$ is limited by the background noise farther than 1\,$''$, consistent with the MIRI commissioning results \citep{boccaletti_jwstmiri_2022}.
The profiles in Fig. \ref{fig:contrast_curve_best}
still show perceptible "bumps," due to stellar diffraction residuals.
The F1065C filter provides poorer contrast than the F1140C at separation closer to 1.5$''$,
which may be due to the fact that the library contains fewer references in the F1065C filter.
The gain at F1550C in using a reference library  is lower than for the two other filters because the background noise is higher,
so the diffraction residuals are not a hard limit even with fewer frames than for F1140C. Moreover, the library features heterogeneous exposures times, hence the shortest exposures drive the overall background noise.
In addition, we also tested different subsets of the reference library and various algorithms to suppress the diffraction pattern. These include the linear optimization of the reference observations, in the same spirit as the algorithm Locally Optimized Combination of Images 
\citep[LOCI,][]{lafreniere_new_2007}, but with a single optimization zone: a ring centered at the coronagraph center from 0.3$''$ to 8$''$.
The companion was detected, regardless of the reference observations used to subtract stellar diffraction, as shown in the appendices \footnote{Appendices are available on \href{https://doi.org/10.5281/zenodo.14517224}{zenodo.14517224}}.
The $S/N$ values for planet detection are indicated in the appendix, 
with the noise measured as the standard deviation of the flux in a ring of 4\,$\lambda$/D ($\sim$ 8 pixels) at the separation of the planet. 
The classical PCA provided the highest $S/N$, but this method has the strongest residuals in the center of the images.
The corresponding contrast curves measured with the different subtraction methods are presented in the appendix.
The method described above using the entire library and the 4Q-PCA method provides the largest contrast.

\section{Atmospheric characterization}
\label{sec:atm_charact}

\subsection{Extraction of the photometry}
\label{sec:photom}

The photometry was extracted as described in \cite{boccaletti_imaging_2024} and \cite{malin_unveiling_2024}.
We use both MJy/sr calibrated images \texttt{$\_$cal} and \texttt{$\_$rates} images, to which we applied a photometric calibration based on the contrast with the host star.
The website \texttt{whereistheplanet} 
\citep{wang_whereistheplanet_2021} 
was used to provide an estimate of the position at the observation date, based on past observations.
We use \texttt{WebbPSF}
\citep{perrin_updated_2014} to simulate the planet PSF models, taking into account the appropriate filter and mask configurations for MIRI coronagraphs.
The planet's position relative to the 4QPM axis was taken into account by specifying its position in the detector coordinates.
The residuals between the PSF model and the data are minimized using the Nelder-Mead algorithm \citep{nelder_simplex_1965}.
The position of the PSF model is also optimized with two free parameters, which shift the model to x- and y-positions.
In the case of PCA-based stellar subtraction methods, the planet's photometry may be biased. Consequently, we modeled the planet's PSF before applying the PCA algorithm and we then minimized the residuals.
This reduces the uncertainties in planet photometry measured on images obtained with the various stellar subtraction methods.
The best-model PSF is presented in the appendix. 
The flux is extracted on the best-fit PSF model image for each filter.
The attenuation due to the coronagraph mask is evaluated with \texttt{WebbPSF} simulations:
we measured the ratio between two simulated PSFs, the first at the planet's position (within the detector frame) and the second at a position unaffected by coronagraph attenuation.
The transmission at GJ\,504\,b's position is 0.84 in  F1065C, 0.93 in  F1140C, and 0.77 in  F1550C. 
Due to different position angles of the telescope, the companion is slightly closer to the transition of the 4QPM axis for the observation at F1065C, justifying the fact that the attenuation is more important than at F1140C.
Each flux value is divided by this attenuation factor in order to recover the emitted flux of the object.
There are two main sources of uncertainties on the flux extracted for the companion.
First, the stellar subtraction can have an impact; thus, we repeated the same procedure for each method of stellar subtraction (with the entire library, as shown in the appendix) 
We obtain consistent values with a variation of
0.8, 1.3, and 6.9\,\%
in the F1065C, F1140C, and F1550C filters, respectively, .
This uncertainty is labeled $\sigma_{stellar sub}$ in the following.
Secondly, the PSF normalization from DN to (W/m$^2$/$\mu$m) is another source of uncertainty, which has not yet been documented.
Therefore, we compared these photometric values with those derived with a contrast approach.
First, we used the method from \cite{boccaletti_imaging_2024}, based on a contrast measurement and estimation of the stellar flux using target acquisition observations.
We added the method based on simulation with \texttt{WebbPSF} to estimate the stellar flux as in \cite{malin_unveiling_2024}.
The different results obtained for the flux measurement are presented in the appendix. 
All methods provided consistent values, but the PSF normalization to obtain physical flux units remains the main source of uncertainty, with a variation of 
4.1, 12.6, and 10.7\,\%
for each of the three filters, respectively.
We refer to this uncertainty value as $\sigma_{PSF norm}$.
Finally, we measured the standard deviation of the residual stellar flux in a ring at the planet separation, to check whether the remaining stellar diffraction may be an additional source of uncertainty in the flux measurement.
These standard deviations are on the order of 0.1\,\% of the flux of the planet in each filter and can consequently be neglected.
The final photometric flux values are indicated in Table \ref{tab:pos_flux}.
They correspond to the averaged flux on the three normalization methods, measured using images from Fig. \ref{fig:image_sub}, which provide the highest $S/N$ values for the detection of GJ\,504\,b.
We took the final uncertainty as the quadratic sum of the two main sources of uncertainties (Eq. \ref{eq:uncertainty}).
\begin{equation}
    \centering
    \sigma_{flux} = \sqrt{\sigma_{stellar sub}^2 + \sigma_{PSF norm}^2}
    \label{eq:uncertainty}
.\end{equation}
The dispersion between the flux values obtained by the different methods is larger in the F1140C filter than in the F1065C, even though the detection occurs with a higher $S/N$ at this wavelength.
\begin{table}[h!]
    \centering
    \caption{Measured photometry for the planet GJ\,504\,b.} 
    \begin{tabular}{c| c}
    \hline
    \hline
    Filters & Flux $\pm$ $\sigma_{flux}$ (W/m$^2$/$\mu$m)\\
    \hline
    F1065C & (3.97 $\pm$ 0.16) $\times$ 10$^{-18}$\\
    F1140C & (4.62 $\pm$ 0.58) $\times$ 10$^{-18}$\\
    F1550C & (1.53 $\pm$ 0.19) $\times$ 10$^{-18}$\\
    \end{tabular}
    \label{tab:pos_flux}
\end{table}

Based on the best fit PSF model, we determined the astrometry relative to the center of the coronagraphic mask to be ($\Delta$RA, $\Delta$DEC) = (-1.75 $\pm$ 0.02$''$, 1.77 $\pm$ 0.01$''$), averaged over the three filters.
The uncertainties correspond to the dispersion of the measurement over the three filters.
This is consistent with previous predictions within the uncertainties and corresponds to a separation of 2.48$\pm$\,0.02$''$ (i.e., 44.4 $\pm$ 0.7\,au at 17.56\,pc).

\subsection{Atmospheric characterization of the planet}
\begin{figure*}[h!]
    \centering
    \includegraphics[width=13cm]{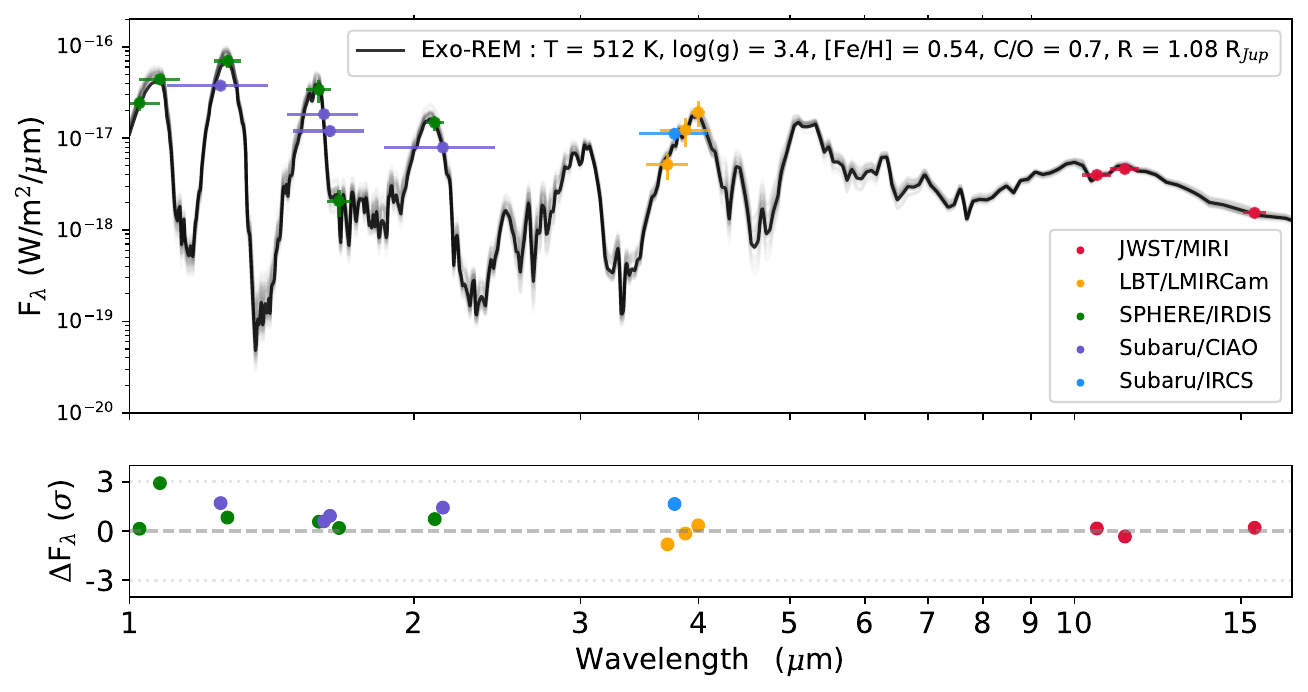}
    \includegraphics[width=5cm]{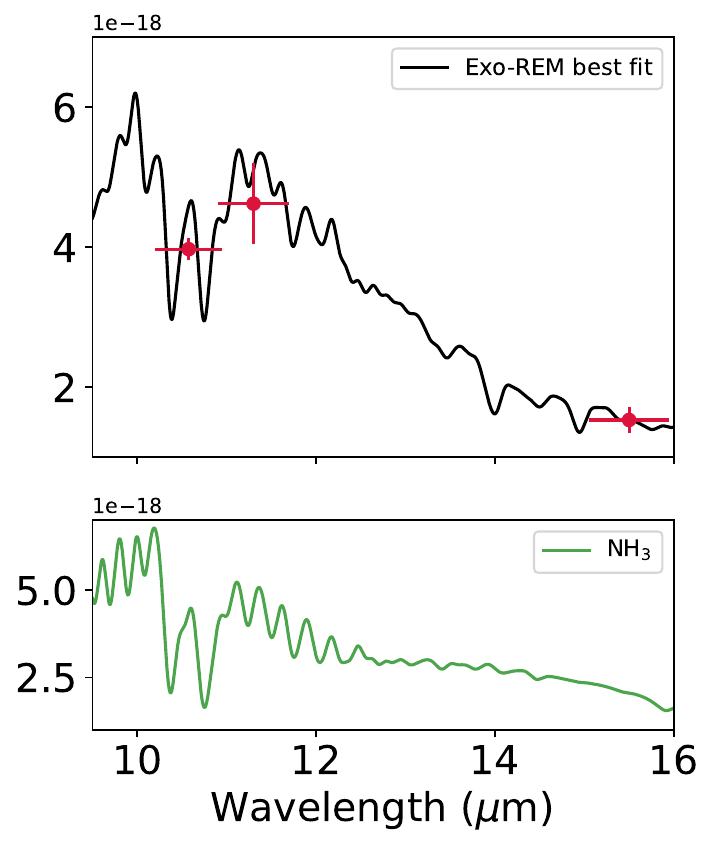}
    \caption{\texttt{Exo-REM} best-fit model (black curve) and the lighter black models come from the posterior distribution from the best fit.
    The points represent the photometric values from VLT/SPHERE (green), Suburu/HiCIAO (purple), LBTI/LMIRcam (orange), and Subaru/ICRS (blue). The MIRI photometry is shown in red.
    Photometric points are plotted at the filter's central wavelength, with the width indicated.
    The bottom plot represent the residuals between the photometric points and the best fits models: all point agree within a 3\,$\sigma$ precision.
    Right: Zoom on the MIRI data and the NH$_3$ absorption, fluxes values are displayed with a linear scale.}
    \label{fig:best_fit_param}
\end{figure*}

We used the self-consistent 1D atmospheric model \texttt{Exo-REM}, developed to simulate the atmosphere of young giant planets \citep{baudino_interpreting_2015} and to understand the L--T transition \citep{charnay_self-consistent_2018}. 
The grid parameter ranges are described in Table \ref{tab:param_models_atm}.
\texttt{Exo-REM} includes a cloud model that takes into account the microphysics (iron and silicate clouds with supersaturation parameter S=0.003). Disequilibrium chemistry is also included in the model.
The sources of opacity include collision-induced absorption of H$_2$--H$_2$, H$_2$--He, H$_2$O--H$_2$O, and H$_2$O-N$_2$, the ro-vibrational bands of molecules (H$_2$O, CH$_4$, CO, CO$_2$, NH$_3$, PH$_3$, TiO, VO, H$_2$S, HCN, and FeH). 
Line lists are given in \cite{blain_1d_2021}.
The NIR photometric values were taken from the literature,  coming from ground-based instruments: Subaru/CIAO \citep{kuzuhara_direct_2013, janson_direct_2013}, 
LBTI/LMIRcam \citep{skemer_leech_2016}, 
and VLT/SPHERE \citep{bonnefoy_gj_2018}.
The points in Fig. \ref{fig:best_fit_param}. represent the photometry values colored-coded by instrument.
No additional scaling factors are used in between instruments.
\begin{table}[H]
    \centering
    \caption{Parameters of the atmospheric grids \texttt{Exo-REM}.}
    \begin{tabular}{c | c c} 
        \hline
        \hline
         Parameters & Range & Step \\
        \hline
        Temperature (K)  & 400 -- 2000 &  50  \\              
        Surface gravity $\mathrm{Log} g$ &  3.0 -- 5.0 & 0.5 \\ 
        C/O              &  0.1 -- 0.8 & 0.05 \\
        Metallicity  & -0.5 -- 1 & 0.5 \\
        \end{tabular}
    \label{tab:param_models_atm}
\end{table}

We used a forward-modeling process, using the python package \texttt{species}\footnote{\texttt{species}: \href{https://species.readthedocs.io/en/latest/}{species.readthedocs.io}} \citep{stolker_species_2023} to measure the atmospheric parameters and their posterior distributions.
We ran the Bayesian analysis with 5000 live points, without prior constraints on the atmospheric parameters.
We ran the analysis twice, once with only the NIR photometry and then again, adding the MIRI photometric points.
The best-fit parameters obtained are summarized in Table \ref{tab:best_fit_param}.
The best-fit model using NIR and MIR photometry is shown in Fig. \ref{fig:best_fit_param}.
The posterior distributions for each parameter are presented in the appendix. 
\begin{table}[h]
    \centering
    \caption{Best-fit atmospheric parameters}
    \begin{tabular}{c|c c}
    \hline
    \hline
    Parameters  &  NIR only & NIR and MIR \\
    \hline
    T$_{\rm eff}$ (K) &  509$^{+13}_{-20}$ & 512$^{+10}_{-10}$ \\
    $\mathrm{log} g$ & 3.42$^{+0.41}_{-0.27}$ & 3.45$^{+0.35}_{-0.25}$\\
    Metallicity  & 0.52$^{+0.11}_{-0.12}$ & 0.54$^{+0.09}_{-0.11}$\\
    C/O  & 0.70$^{+0.06}_{-0.07}$ & 0.70$^{+0.06}_{-0.07}$\\
    Radius (R$_{\rm Jup}$)  & 1.13$^{+0.16}_{-0.14}$ & 1.08$^{+0.04}_{-0.03}$\\
    \hline
    Luminosity $\mathrm{log}$(L/L$_{\sun}$) & -6.09$^{+0.07}_{-0.07}$ & -6.12$^{+0.02}_{-0.02}$\\
    Mass (M$_{\rm Jup}$) & 1.3$^{+1.8}_{-0.5}$ & 1.0$^{+1.8}_{-0.3}$\\
    \end{tabular}
    \label{tab:best_fit_param}
\end{table}

The atmospheric parameters measured by adding MIRI photometric points are consistent to the 1\,$\sigma$ level
with the atmospheric parameters previously estimated, based solely on near-infrared data.
The mass value is derived from the surface gravity and radius values; and the luminosity is measured from the temperature and radius.
Expanding the wavelength range reduces uncertainties in the radius measurement by at least a factor 3, consequently enhancing the precision of the logarithmic measure of luminosity 
by a factor $\sim$ 3.5.
These parameters are evaluated independently of the assumptions about the age of the system and the evolution models.
The mass values seem to be more consistent with the young age hypothesis; however, it relies on the surface gravity measurement, which is not confidently reliable with only photometric points.
Indeed, the surface gravity is embedded in the shape of the lines rather than on the continuum.\\

NH$_3$ absorption explains the lower flux at F1065C than at F1140C.
We generated a second \texttt{Exo-REM} atmospheric grid based on the best-fit parameters and varying the volume mixing ratio (vmr) of NH$_{3}$ from 5$\cdot10^{-10}$ to 1$\cdot10^{-4}$.
Following \cite{danielski_atmospheric_2018},
we measured the detection level of NH$_3$ by comparing it to the model with the lowest NH$_3$ abundance, set at 5$\cdot10^{-10}$, which has a negligible effect on the spectrum.
It is expressed as%
\begin{equation}
    \centering
    \label{eq:snr_nh3}
    S/N_{(NH_3)} = \frac{{F_{no NH_3, F1065C} - F_{obs, F1065C}}}{\sigma_{tot}}
,\end{equation}
where $\sigma_{tot}$ is the quadratic sum of the relative uncertainties on the measured flux,
\begin{equation}
    \label{eq:sigma_tot}
    \sigma_{tot} = \sqrt{\sigma_{F1065C}^2 + \sigma_{F1140C}^2} 
.\end{equation}

Comparing the flux difference between F1065C and F1140C to the photometric errors (derived in Sect. \ref{sec:photom}) yields a result of $S/N_{(NH_3)}=3\sigma$. However, these uncertainties correspond to an absolute photometric precision; hence, the outcome for NH$_3$ is quite conservative and must be taken as a lower limit. 
Instead, given it is a relative measurement, a more reliable estimate of the NH$_3$ detection level should be derived from each photometric method independently. 
This means zeroing the term $\sigma_{PSFnorm}^2$ in 
Eq. \ref{eq:uncertainty}.
Taking this precaution provides a more reliable estimate at 12.5\,$\sigma$.
Finally, we calculate the $\chi^2$ values for each model in this second atmospheric grid. 
Models with a vmr of NH$_3$ in the range -5.37 and -5.22 (in log$_{10}$) fall within the 1\,$\sigma$ confidence level of the $\chi^2$ minimisation
(see the figure in the appendix).

\section{Discussion}
\label{sec:discussion}

\subsection{Stellar subtraction}
To remove the stellar diffraction,
we reconstructed an optimized reference image subtracted from the data using traditional algorithms previously developed for observations with ground-based instruments (such as PCA and linear combination of references images).
These algorithms are adapted for MIRI coronagraph data.
Moreover, we used individual sets of references from a specific program and then the entire reference library.
We found that using a single set of reference is not effective enough to subtract the stellar diffraction pattern from coronagraphic images, at least in the particular case of GJ\,504 observations.
We note that the bright residuals that could be assimilated as point sources are highly dependent on the reference dataset and the stellar diffraction subtraction methods used.
Hence, they cannot be considered as real new point sources.

The reference star from the GTO\,1413 program is of a spectral type and magnitude similar to GJ\,504\,A and it was observed a few days after the GJ\,504 system.
It represented an optimistic dataset to provide a good stellar subtraction.
However, a small wavefront drift occurred between these two observations and the wavefront error increased by $\sim$ 6 nm (measures available with \texttt{WebbPSF}), which could have had an impact on the quality of this data set.
The quality of stellar subtraction is improved when we increase the number of datasets in the reference library, thereby achieving more favorable contrasts.
%
Previous works conducted within this GTO\,1277 program \citep[systems HR\,8799 and HD\,95086;][]{boccaletti_imaging_2024, malin_unveiling_2024} indicated that the use of a reference library did not yield an improved stellar subtraction.
For these observations, dedicated reference stars were observed together with the scientific target.
Furthermore, the library used was built with fewer references available, coming only from the ERS, GTO and commissioning programs.
We tested our updated library, containing all GTO and GO programs available (cycle 1 and part of cycle 2) for these two datasets, but it does not provide improved results.
We conclude that obtaining reference observations captured in the same sequence as the target observations remains the optimal method for mitigating stellar diffraction in coronagraphic images, especially for closer-in planets (such as the planets from the systems HD\,95086 and HR\,8799).
However, the accumulation of a larger reference library in the next few years could provide better results in the future, allowing for the capture and removal of diffraction residuals at separations shorter than 1$''$.

Furthermore, the PCA analysis provides better subtraction \citep[such as for HIP\,65426 system,][]{carter_jwst_2023} for GJ\,504, in contrast to the HR\,8799 and HD\,95086 systems
for which the optimized linear combination of reference yields a better subtraction.
This is likely due to the fact that the systems GJ\,504 and HIP\,65426 do not contain a warm inner disk (or multiple bright planets and prominent background objects), which prevents a straightforward PCA-based stellar subtraction.
In conclusion, the best algorithm for subtracting the stellar diffraction in MIRI data strongly depends on the architecture of the system itself.

The asymmetries in the stellar residuals visible in the coronagraphic image due to the 4QPM led us to apply the PCA in each quadrant independently.
Future algorithms adapted specifically to the 4QPM coronagraph could improve the stellar subtraction.
In the case of the GJ\,504 system, using the full library provides all the diversity available.
We note that we did not get any improvement in the performance when we selected samples from the library.
For more challenging observations (fainter targets or closer separations), we could achieve a better performance by evaluating the choice of references in the library, avoiding those containing an inner disk, and optimizing the regions of the field of view where the residuals should be minimized.
%


\subsection{Spectral characterization}
\label{subsec:atm_caract}
We present the detection of NH$_3$ in the atmosphere of a directly imaged planetary-mass companion at 12.5\,$\sigma$
using the relative measurement of the flux between the two filters F1065C centered on the absorption of ammonia and F1140C in the continuum of the spectrum.
Using a conservative measurement of the flux and uncertainties that takes several methods into account, we estimated a lower limit of detection of 3$\sigma$.
Even though it has been detected in many T-type isolated brown dwarfs \citep[for example,][]{suarez_ultracool_2022},
this molecule has been previously inferred in the atmosphere of only one directly imaged planet  \citep[detection at 2.7\,$\sigma$ for 51\,Eri\,b using retrieval analysis with its NIR data by][]{whiteford_retrieval_2023}.
Using pre-computed \texttt{ExoREM} grids and varying the volume mixing ratio of NH$_3$, we measured a vmr of 10$^{-5.3 \pm 0.07}$ for GJ\,504\,b.
For comparison, this is at least an order of magnitude lower than in Jupiter's atmosphere \citep{taylor_composition_2004}.

\begin{figure}
    \centering
    \includegraphics[width=0.8\linewidth]{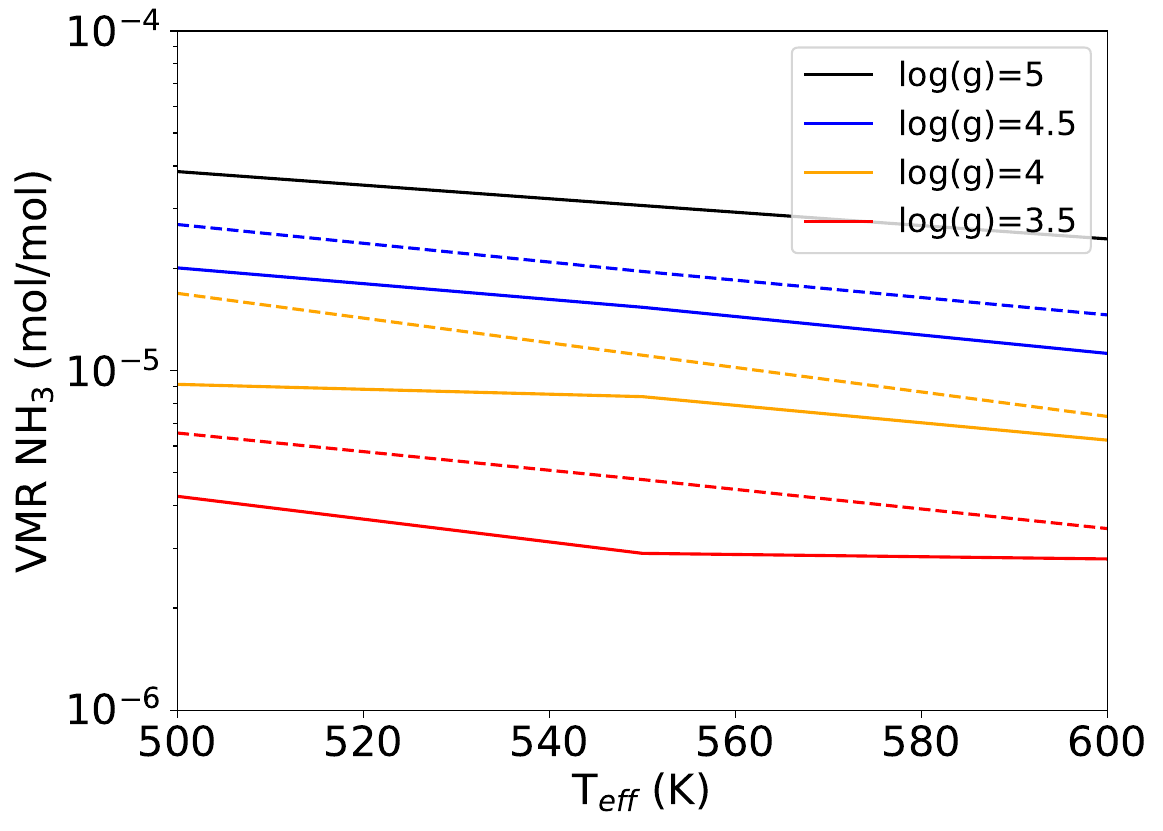}
    \caption{Abundance of the NH$_3$ according to \texttt{Exo-REM} models as a function of the $T_{eff}$ for several values of $\mathrm{log} g$. 
    The dashed lines represent the expectation for an atmosphere with a super-solar metallicity, and the plain line with a solar metallicity.}
    \label{fig:nh3_exorem}
\end{figure}

The NH$_3$ is sensitive to gravity (higher NH$_3$ abundance with higher surface gravity, as shown in Fig. \ref{fig:nh3_exorem}) and unlike the CH$_4$/CO ratio, its ratio NH$_3$/N$_2$ is insensitive to mixing, making this molecule an interesting proxy for gravity \citep{zahnle_methane_2014}.
In the case of a planetary mass object for which T$_{\rm eff}$ and the radius can be measured more accurately, the measurement of the abundance of NH$_3$ could therefore provide an additional estimate of the mass, since mass is directly linked to radius and surface gravity.
The measured value of NH$_3$ abundance for GJ\,504\,b is more consistent with an object of a surface gravity $\mathrm{log} g$ < 4, according to \texttt{Exo-REM} models (Fig. \ref{fig:nh3_exorem}). 
As a result, GJ\,504\,b photometry appears more consistent with a planetary-mass object rather than a brown dwarf. 
The presence of clouds in the atmosphere may affect this measurement, albeit to a lesser extent than metallicity.
The detection of NH$_3$ is also promising to allow for the measurement of the isotopic ratio $^{14}$N/$^{15}$N, as done for isolated Y-type brown dwarfs with MIRI/MRS spectra \citep{barrado_15nh3_2023, kuhnle_water_2024_a}.
%
Young stars and consequently their planets should be more strongly enriched in the $^{15}$N isotope 
\citep{adande_millimeter-wave_2012}.
Determining the $^{14}$N/$^{15}$N isotopic ratio, the C/N ratio \citep{turrini_tracing_2021, pacetti_chemical_2022}, and C/O ratio
\citep[e.g.,][]{oberg_effects_2011, madhusudhan_exoplanetary_2019}
can provide important constraints on the formation location and pathway of GJ\,504\,b.\\

\begin{figure}[h!]
    \centering
    \includegraphics[width=9cm]{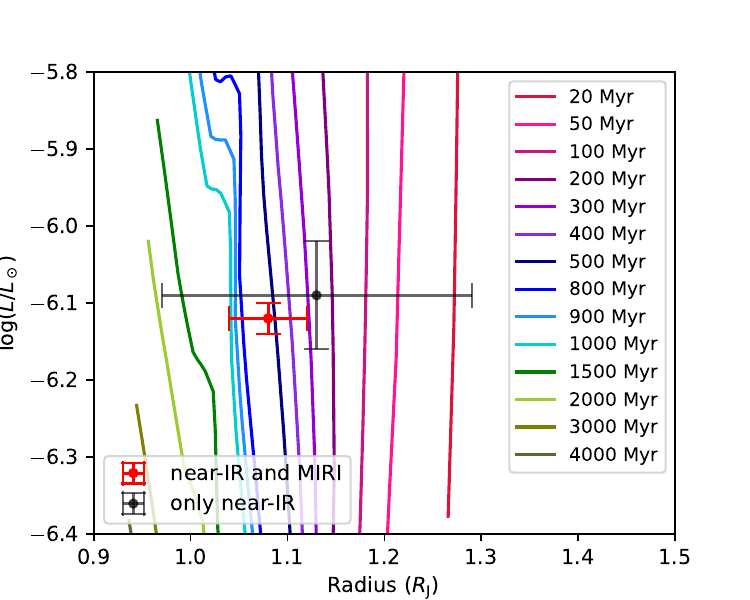}
    \includegraphics[width=9cm]{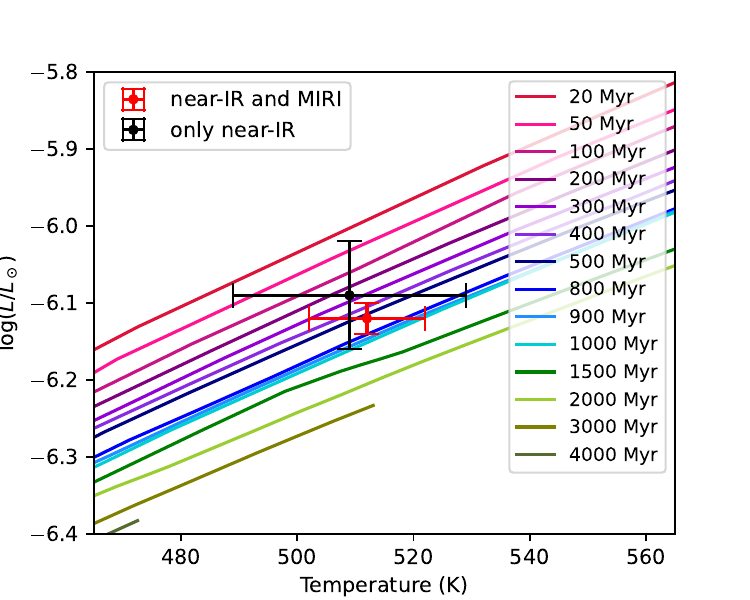}
    \caption{Isochrones from 20 Myr to 4 Gyr: evolution of the luminosity of a planet (log scale and relative to the Sun's luminosity) as a function of the radius (top panel) or the temperature (bottom panel) of a planet according to \texttt{ATMO} evolutionary models.}
    \label{fig:evol_modelsl}
\end{figure}

The atmospheric parameters measured (temperature, radius, and luminosity) are compared to \texttt{ATMO} evolutionary models that include disequilibrium chemistry \citep[][]{phillips_new_2020} in Fig. \ref{fig:evol_modelsl}.
A smaller radius is measured when the MIR photometric points are added, together with lower uncertainties 
(red points, in comparison to the black points, Fig. \ref{fig:evol_modelsl}).
The comparison of luminosity and measured radius corresponds to isochrones between 400\,Myr and 1\,Gyr (at 1\,$\sigma$).
Compared with the Sonora evolution models \citep{marley_sonora_2021}, we find that the radius and effective temperature measured from the atmospheric fit correspond to isochrones slightly older, from $\sim$ 500\,Myr to 1.5\,Gyr.
The same trend is observed when the measured values are compared with the effective temperature expected from evolution models, rather than with the radius value.
The MIR photometry seems to place GJ\,504\,b within the older age range.
However, this is insufficient to confirm the nature of GJ\,504\,b, as this range of stellar ages translates into a wide range of masses. 
Indeed, for the isochrones between 400\,Myr to 1\,Gyr, the masses from $\sim$ 1 to 17 M$_{\rm Jup}$ are consistent with the measured radius.

The mass measured from the surface gravity of the best-fit atmospheric model corresponds to the low-mass hypothesis.
One can argue that the $\mathrm{log} g$ value is not reliable when fitting only a few photometric points, as this parameter is measured with higher uncertainties.
At MIR wavelengths, surface gravity influences the shape of the spectral lines rather than the spectral continuum.
Finally, the radius value might still be inconsistent with evolutionary models. 
Even with the addition of MIR information, atmospheric measurements still provide a slightly smaller radius than that predicted by the evolution models. 
Therefore, confirming the nature of GJ\,504\,b based on radius measurement may not be reliable.
A deeper atmospheric analysis including different atmospheric models is outside the scope of this study, but should be carried out in the future with the coming JWST observation of GJ\,504\,b providing high $S/N$ spectra at both NIR and MIR wavelengths.


\section{Conclusion and perspective}
\label{sec:conclusion}
In this paper, we present our analysis of the MIRI coronagraphic images of one of the coldest directly imaged planetary-mass companion to date. Our findings are as follows.
\begin{itemize}
    \item The stellar diffraction was subtracted from the coronagraph image using a reference library built on previous JWST observations.
    For the first time, we  used a large library with all the observations available up to June 2024.
    This provides better performance than the previous library \citep[used in][]{boccaletti_imaging_2024, malin_unveiling_2024}, which only contained ERS and early GTO observations
    Even though this method does not perfectly remove the stellar residuals at shorter separation, it holds promise for achieving a better contrast when no reference observations are available. 
    \item We confidently detected  the presence of NH$_3$ in the atmosphere of a late T-type planetary mass companion.
    \item 
    The atmospheric parameters were constrained with lower uncertainties, owing to photometric values measured at MIR wavelengths.
    The luminosity and radius measurement are more accurate by at least a factor of 3.
    \item The addition of MIRI photometry enabled the measurement of a smaller radius of R = 1.08 R$_{\rm Jup}$.
    This value is in agreement with the isochrones from 400\,Myr to 1\,Gyr, according to evolutionary models.
    \item Although the MIR photometry is still not sufficient to narrow down the mass and confirm the planetary nature of GJ\,504\,b; these MIRI observations show the first MIR data and are valuable for forthcoming spectroscopic observations of this object.
\end{itemize}
The GJ\,504 system has recently been observed with both integral-field spectrographs of JWST (NIRSpec/IFU, GTO 2778, PI: M. Perrin and MIRI/MRS, GO 3647, PI: P. Patapis).
Furthermore, recent observations from VLT/CRIRES (PI: F. Kiefer) will provide high-resolution data at NIR wavelengths.
The MIRI photometry  presented here offers new constraints on this T-dwarf planetary mass companion, which will prove useful in analyses of the forthcoming datasets.
The wealth of spectroscopic data coming will allow for an improved characterization of its atmosphere, while confirming its nature and to allowing us to better understand its formation and evolution pathway.


\begin{acknowledgement}
This work is based on observations made with the NASA/ESA/CSA James Webb Space Telescope. The data were obtained from the Mikulski Archive for Space Telescopes at the Space Telescope Science Institute, which is operated by the Association of Universities for Research in Astronomy, Inc., under NASA contract NAS 5-03127 for JWST. These observations are associated with program \#1277.
Part of this work was carried out at the Jet Propulsion Laboratory, California Institute of Technology, under contrast with NASA (80NM0018D0004).
This publication makes use of VOSA, developed under the Spanish Virtual Observatory (https://svo.cab.inta-csic.es) project funded by MCIN/AEI/10.13039/501100011033/ through grant PID2020-112949GB-I00. 
VOSA has been partially updated by using funding from the European Union's Horizon 2020 Research and Innovation Programme, under Grant Agreement nº 776403 (EXOPLANETS-A)
This work has benefitted from The UltracoolSheet at http://bit.ly/UltracoolSheet, maintained by Will Best, Trent Dupuy, Michael Liu, Aniket Sanghi, Rob Siverd, and Zhoujian Zhang, and developed from compilations by  \cite{dupuy_hawaii_2012}, \cite{dupuy_distances_2013}, \cite{deacon_wide_2014}, \cite{liu_hawaii_2016}, \cite{best_photometry_2018}, \cite{best_volume-limited_2021}, \cite{sanghi_hawaii_2023}, and \cite{schneider_astrometry_2023}.
This work has made use of data from the European Space Agency (ESA) mission
{\it Gaia} (\url{https://www.cosmos.esa.int/gaia}), processed by the {\it Gaia}
Data Processing and Analysis Consortium (DPAC,
\url{https://www.cosmos.esa.int/web/gaia/dpac/consortium}). Funding for the DPAC
has been provided by national institutions, in particular the institutions
participating in the {\it Gaia} Multilateral Agreement.
This research has made use of the VizieR catalogue access tool, CDS, Strasbourg, France (DOI : 10.26093/cds/vizier). The original description of the VizieR service was published in 2000, A\&AS 143, 23
Software: python, astropy, numpy, scipy  matplotlib.
M.M., A.B., P.-O.L, C.C. acknowledges funding support by CNES.
B.V. thanks the European Space Agency (ESA) and the Belgian Federal Science Policy Office (BELSPO) for their support in the framework of the PRODEX Programme.
O.A. is a Senior Research Associate of the Fonds de la Recherche Scientifique – FNRS. OA thanks the European Space Agency (ESA) and the Belgian Federal Science Policy Office (BELSPO) for their support in the framework of the PRODEX Programme.
D.B. is supported by Spanish MCIN/AEI/10.13039/501100011033 grant PID2019-107061GB-C61 and No. MDM-2017-0737. 
L.D. acknowledges funding from the KU Leuven Interdisciplinary Grant (IDN/19/028), the European Union H2020-MSCA-ITN-2019 under Grant no. 860470 (CHAMELEON) and the FWO research grant G086217N.
MPIA acknowledges support from the Federal Ministry of Economy (BMWi) through the German Space Agency (DLR).
J.P. acknowledges financial support from the UK Science and Technology Facilities Council, and the UK Space Agency.
G.O. acknowledge support from the Swedish National Space Board and the Knut and Alice Wallenberg Foundation.
P.P. thanks the Swiss National Science Foundation (SNSF) for financial support under grant number 200020\_200399.
P.R. thanks the European Space Agency (ESA) and the Belgian Federal Science Policy Office (BELSPO) for their support in the framework of the PRODEX Programme.
E.C acknowledges funding by the European Union (ERC, ESCAPE, project No 101044152). Views and opinions expressed are however those of the author(s) only and do not necessarily reflect those of the European Union or the European Research Council Executive Agency. Neither the European Union nor the granting authority can be held responsible for them.
C.D. acknowledges support from the INAF initiative “IAF Astronomy Fellowships in Italy”, grant name \textit{GExoLife}.
E. v D acknowledges support from A-ERC grant 101019751 MOLDISK.
The Cosmic Dawn Center (DAWN) is funded by the Danish National Research Foundation under grant No. 140. TRG is grateful for support from the Carlsberg Foundation via grant No. CF20-0534.
T.P.R acknowledges support from the ERC 743029 EASY. 
Support from SNSA is acknowledged.
\end{acknowledgement}

\bibliographystyle{aa} 
\bibliography{main} 

\begin{thebibliography}{56}
\expandafter\ifx\csname natexlab\endcsname\relax\def\natexlab#1{#1}\fi

\bibitem[{Adande \& Ziurys(2012)}]{adande_millimeter-wave_2012}
Adande, G.~R. \& Ziurys, L.~M. 2012, ApJ, 744, 194

\bibitem[{Anderson {et~al.}(2010)Anderson, Reiners, \& Solanki}]{anderson_detectability_2010}
Anderson, R.~I., Reiners, A., \& Solanki, S.~K. 2010, A \& A, 522, A81

\bibitem[{Barrado {et~al.}(2023)Barrado, Mollière, Patapis, Min, Tremblin, Martinez, Whiteford, Vasist, Argyriou, Samland, Lagage, Decin, Waters, Henning, Morales-Calderón, Guedel, Vandenbussche, Absil, Baudoz, Boccaletti, Bouwman, Cossou, Coulais, Crouzet, Gastaud, Glasse, Glauser, Kamp, Kendrew, Krause, Lahuis, Mueller, Olofsson, Pye, Rouan, Royer, Scheithauer, Waldmann, Colina, van Dishoeck, Ray, Östlin, \& Wright}]{barrado_15nh3_2023}
Barrado, D., Mollière, P., Patapis, P., {et~al.} 2023, Nature, 624, 263

\bibitem[{Baudino {et~al.}(2015)Baudino, Bézard, Boccaletti, Bonnefoy, Lagrange, \& Galicher}]{baudino_interpreting_2015}
Baudino, J.-L., Bézard, B., Boccaletti, A., {et~al.} 2015, A\&A, 582, A83

\bibitem[{Best {et~al.}(2024)Best, Dupuy, Liu, Sanghi, Siverd, \& Zhang}]{best_ultracoolsheet_2024}
Best, W. M.~J., Dupuy, T.~J., Liu, M.~C., {et~al.} 2024, Zenodo

\bibitem[{Best {et~al.}(2021)Best, Liu, Magnier, \& Dupuy}]{best_volume-limited_2021}
Best, W. M.~J., Liu, M.~C., Magnier, E.~A., \& Dupuy, T.~J. 2021, The Astronomical Journal, 161, 42

\bibitem[{Best {et~al.}(2018)Best, Magnier, Liu, Aller, Zhang, Burgett, Chambers, Draper, Flewelling, Kaiser, Kudritzki, Metcalfe, Tonry, Wainscoat, \& Waters}]{best_photometry_2018}
Best, W. M.~J., Magnier, E.~A., Liu, M.~C., {et~al.} 2018, The Astrophysical Journal Supplement Series, 234, 1

\bibitem[{Blain {et~al.}(2021)Blain, Charnay, \& Bézard}]{blain_1d_2021}
Blain, D., Charnay, B., \& Bézard, B. 2021, A\&A, 646, A15

\bibitem[{Boccaletti {et~al.}(2022)Boccaletti, Cossou, Baudoz, Lagage, Dicken, Glasse, Hines, Aguilar, Detre, Nickson, Noriega-Crespo, G\'asp\'ar, Labiano, Stark, Rouan, Reess, Wright, Rieke, Garcia~Marin, Lajoie, Girard, Perrin, Soummer, \& Pueyo}]{boccaletti_jwstmiri_2022}
Boccaletti, A., Cossou, C., Baudoz, P., {et~al.} 2022, A\&A, 667, A165

\bibitem[{Boccaletti {et~al.}(2024)Boccaletti, Mâlin, Baudoz, Tremblin, Perrot, Rouan, Lagage, Whiteford, Mollière, Waters, Henning, Decin, Güdel, Vandenbussche, Absil, Argyriou, Bouwman, Cossou, Coulais, Gastaud, Glasse, Glauser, Kamp, Kendrew, Krause, Lahuis, Mueller, Olofsson, Patapis, Pye, Royer, Serabyn, Scheithauer, Colina, Van~Dishoeck, Ostlin, Ray, \& Wright}]{boccaletti_imaging_2024}
Boccaletti, A., Mâlin, M., Baudoz, P., {et~al.} 2024, A\&A, 686, A33

\bibitem[{Bodenheimer(2000)}]{bodenheimer_models_2000}
Bodenheimer, P. 2000, Icarus, 143, 2

\bibitem[{Bonnefoy {et~al.}(2018)Bonnefoy, Perraut, Lagrange, Delorme, Vigan, Line, Rodet, Ginski, Mourard, Marleau, Samland, Tremblin, Ligi, Cantalloube, Mollière, Charnay, Kuzuhara, Janson, Morley, Homeier, D’Orazi, Klahr, Mordasini, Lavie, Baudino, Beust, Peretti, Musso~Bartucci, Mesa, Bézard, Boccaletti, Galicher, Hagelberg, Desidera, Biller, Maire, Allard, Borgniet, Lannier, Meunier, Desort, Alecian, Chauvin, Langlois, Henning, Mugnier, Mouillet, Gratton, Brandt, Mc~Elwain, Beuzit, Tamura, Hori, Brandner, Buenzli, Cheetham, Cudel, Feldt, Kasper, Keppler, Kopytova, Meyer, Perrot, Rouan, Salter, Schmidt, Sissa, Zurlo, Wildi, Blanchard, De~Caprio, Delboulbé, Maurel, Moulin, Pavlov, Rabou, Ramos, Roelfsema, Rousset, Stadler, Rigal, \& Weber}]{bonnefoy_gj_2018}
Bonnefoy, M., Perraut, K., Lagrange, A.-M., {et~al.} 2018, A\&A, 618, A63

\bibitem[{Bushouse {et~al.}(2023)Bushouse, Eisenhamer, Dencheva, Davies, Morrison, Hodge, Simon, Grumm, Droettboom, Slavich, Sosey, Pauly, Miller, Jedrzejewski, Davis, Crawford, Law, \& Gordon}]{bushouse_jwst_20245}
Bushouse, H., Eisenhamer, J., Dencheva, N., {et~al.} 2023, Zenodo

\bibitem[{Carter {et~al.}(2023)Carter, Hinkley, Kammerer, Skemer, Biller, Leisenring, Millar-Blanchaer, Petrus, Stone, Ward-Duong, Wang, Girard, Hines, Perrin, Pueyo, Balmer, Bonavita, Bonnefoy, Chauvin, Choquet, Christiaens, Danielski, Kennedy, Matthews, Miles, Patapis, Ray, Rickman, Sallum, Stapelfeldt, Whiteford, Zhou, Absil, Boccaletti, Booth, Bowler, Chen, Currie, Fortney, Grady, Greebaum, Henning, Hoch, Janson, Kalas, Kenworthy, Kervella, Kraus, Lagage, Liu, Macintosh, Marino, Marley, Marois, Matthews, Mawet, McElwain, Metchev, Meyer, Molliere, Moran, Morley, Mukherjee, Pantin, Quirrenbach, Rebollido, Ren, Schneider, Vasist, Worthen, Wyatt, Briesemeister, Bryan, Calissendorff, Cantalloube, Cugno, De~Furio, Dupuy, Factor, Faherty, Fitzgerald, Franson, Gonzales, Hood, Howe, Kuzuhara, Lagrange, Lawson, Lazzoni, Lew, Liu, Llop-Sayson, Lloyd, Martinez, Mazoyer, Palma-Bifani, Quanz, Redai, Samland, Schlieder, Tamura, Tan, Uyama, Vigan, Vos, Wagner, Wolff, Ygouf, Zhang, Zhang, \& Zhang}]{carter_jwst_2023}
Carter, A.~L., Hinkley, S., Kammerer, J., {et~al.} 2023, ApJ Letters, 951, L20

\bibitem[{Charnay {et~al.}(2018)Charnay, Bézard, Baudino, Bonnefoy, Boccaletti, \& Galicher}]{charnay_self-consistent_2018}
Charnay, B., Bézard, B., Baudino, J.-L., {et~al.} 2018, ApJ, 854, 172

\bibitem[{Danielski {et~al.}(2018)Danielski, Baudino, Lagage, Boccaletti, Gastaud, Coulais, \& Bézard}]{danielski_atmospheric_2018}
Danielski, C., Baudino, J.-L., Lagage, P.-O., {et~al.} 2018, The Astronomical Journal, 156, 276

\bibitem[{Deacon {et~al.}(2014)Deacon, Liu, Magnier, Aller, Best, Dupuy, Bowler, Mann, Redstone, Burgett, Chambers, Draper, Flewelling, Hodapp, Kaiser, Kudritzki, Morgan, Metcalfe, Price, Tonry, \& Wainscoat}]{deacon_wide_2014}
Deacon, N.~R., Liu, M.~C., Magnier, E.~A., {et~al.} 2014, The Astrophysical Journal, 792, 119

\bibitem[{Di~Mauro {et~al.}(2022)Di~Mauro, Reda, Mathur, Garc\'ia, Buzasi, Corsaro, Benomar, Gonz\'alez~Cuesta, Stassun, Benatti, D’Orazi, Giovannelli, Mesa, \& Nardetto}]{di_mauro_characterization_2022}
Di~Mauro, M.~P., Reda, R., Mathur, S., {et~al.} 2022, ApJ, 940, 93

\bibitem[{Dupuy \& Kraus(2013)}]{dupuy_distances_2013}
Dupuy, T.~J. \& Kraus, A.~L. 2013, Science, 341, 1492

\bibitem[{Dupuy \& Liu(2012)}]{dupuy_hawaii_2012}
Dupuy, T.~J. \& Liu, M.~C. 2012, The Astrophysical Journal Supplement Series, 201, 19

\bibitem[{D’Orazi {et~al.}(2017)D’Orazi, Desidera, Gratton, Lanza, Messina, Andrievsky, Korotin, Benatti, Bonnefoy, Covino, \& Janson}]{dorazi_critical_2017}
D’Orazi, V., Desidera, S., Gratton, R.~G., {et~al.} 2017, A\&A, 598, A19

\bibitem[{Fuhrmann \& Chini(2015)}]{fuhrmann_age_2015}
Fuhrmann, K. \& Chini, R. 2015, ApJ, 806, 163

\bibitem[{Godoy {et~al.}(2024)Godoy, Choquet, Serabyn, Danielski, Stolker, Charnay, Hinkley, Lagage, Ressler, Tremblin, \& Vigan}]{godoy_new_2024}
Godoy, N., Choquet, E., Serabyn, E., {et~al.} 2024, A\&A, 689, A185

\bibitem[{Janson {et~al.}(2013)Janson, Brandt, Kuzuhara, Spiegel, Thalmann, Currie, Bonnefoy, Zimmerman, Sorahana, Kotani, Schlieder, Hashimoto, Kudo, Kusakabe, Abe, Brandner, Carson, Egner, Feldt, Goto, Grady, Guyon, Hayano, Hayashi, Hayashi, Henning, Hodapp, Ishii, Iye, Kandori, Knapp, Kwon, Matsuo, McElwain, Mede, Miyama, Morino, Moro-Mart\'in, Nakagawa, Nishimura, Pyo, Serabyn, Suenaga, Suto, Suzuki, Takahashi, Takami, Takato, Terada, Tomono, Turner, Watanabe, Wisniewski, Yamada, Takami, Usuda, \& Tamura}]{janson_direct_2013}
Janson, M., Brandt, T.~D., Kuzuhara, M., {et~al.} 2013, ApJ, 778, L4

\bibitem[{Kuzuhara {et~al.}(2013)Kuzuhara, Tamura, Kudo, Janson, Kandori, Brandt, Thalmann, Spiegel, Biller, Carson, Hori, Suzuki, Burrows, Henning, Turner, McElwain, Moro-Mart\'in, Suenaga, Takahashi, Kwon, Lucas, Abe, Brandner, Egner, Feldt, Fujiwara, Goto, Grady, Guyon, Hashimoto, Hayano, Hayashi, Hayashi, Hodapp, Ishii, Iye, Knapp, Matsuo, Mayama, Miyama, Morino, Nishikawa, Nishimura, Kotani, Kusakabe, Pyo, Serabyn, Suto, Takami, Takato, Terada, Tomono, Watanabe, Wisniewski, Yamada, Takami, \& Usuda}]{kuzuhara_direct_2013}
Kuzuhara, M., Tamura, M., Kudo, T., {et~al.} 2013, ApJ, 774, 11

\bibitem[{Kühnle {et~al.}(2024)Kühnle, Patapis, Mollière, Tremblin, Matthews, Glauser, Whiteford, Vasist, Absil, Barrado, Min, Lagage, Waters, Guedel, Henning, Vandenbussche, Baudoz, Decin, Pye, Royer, Dishoeck, Östlin, Ray, \& Wright}]{kuhnle_water_2024_a}
Kühnle, H., Patapis, P., Mollière, P., {et~al.} 2024, ArXiv e-prints [\eprint[arXiv]{2410.10933}]

\bibitem[{Lafreniere {et~al.}(2007)Lafreniere, Marois, Doyon, Nadeau, \& Artigau}]{lafreniere_new_2007}
Lafreniere, D., Marois, C., Doyon, R., Nadeau, D., \& Artigau, E. 2007, ApJ, 660, 770

\bibitem[{Lajoie {et~al.}(2016)Lajoie, Soummer, Pueyo, Hines, Nelan, Perrin, Clampin, \& Isaacs}]{lajoie_small-grid_2016}
Lajoie, C.-P., Soummer, R., Pueyo, L., {et~al.} 2016, in {SPIE} {Proceedings}, Edinburgh, United Kingdom, 99045K

\bibitem[{Liu {et~al.}(2016)Liu, Dupuy, \& Allers}]{liu_hawaii_2016}
Liu, M.~C., Dupuy, T.~J., \& Allers, K.~N. 2016, The Astrophysical Journal, 833, 96

\bibitem[{Madhusudhan(2019)}]{madhusudhan_exoplanetary_2019}
Madhusudhan, N. 2019, Annual Review of A\&A, 57, 617

\bibitem[{Marley {et~al.}(2007)Marley, Fortney, Hubickyj, Bodenheimer, \& Lissauer}]{marley_luminosity_2007}
Marley, M.~S., Fortney, J.~J., Hubickyj, O., Bodenheimer, P., \& Lissauer, J.~J. 2007, ApJ, 655, 541

\bibitem[{Marley {et~al.}(2021)Marley, Saumon, Visscher, Lupu, Freedman, Morley, Fortney, Seay, Smith, Teal, \& Wang}]{marley_sonora_2021}
Marley, M.~S., Saumon, D., Visscher, C., {et~al.} 2021, ApJ, 920, 85

\bibitem[{Marston {et~al.}(2018)Marston, Shaw, Forshay, Levay, Mullally, \& Hargis}]{marston_overview_2018}
Marston, A., Shaw, R.~A., Forshay, P., {et~al.} 2018, in Observatory {Operations}: {Strategies}, {Processes}, and {Systems} {VII}, ed. A.~B. Peck, C.~R. Benn, \& R.~L. Seaman (Austin, United States: SPIE), 41

\bibitem[{Matthews {et~al.}(2024)Matthews, Carter, Pathak, Morley, Phillips, P.~M, Feng, Bonse, Boogaard, Burt, Crossfield, Douglas, Henning, Hom, Ko, Kasper, Lagrange, Petit Dit De La~Roche, \& Philipot}]{matthews_temperate_2024}
Matthews, E.~C., Carter, A.~L., Pathak, P., {et~al.} 2024, Nature, 633, 789

\bibitem[{Mollière {et~al.}(2022)Mollière, Molyarova, Bitsch, Henning, Schneider, Kreidberg, Eistrup, Burn, Nasedkin, Semenov, Mordasini, Schlecker, Schwarz, Lacour, Nowak, \& Schulik}]{molliere_interpreting_2022}
Mollière, P., Molyarova, T., Bitsch, B., {et~al.} 2022, ApJ, 934, 74

\bibitem[{Mâlin {et~al.}(2024)Mâlin, Boccaletti, Perrot, Baudoz, Rouan, Lagage, Waters, Güdel, Henning, Vandenbussche, Absil, Barrado, Cossou, Decin, Glauser, Pye, Olofsson, Glasse, Lahuis, Patapis, Royer, Scheithauer, Whiteford, Serabyn, Choquet, Colina, Ostlin, Ray, \& Wright}]{malin_unveiling_2024}
Mâlin, M., Boccaletti, A., Perrot, C., {et~al.} 2024, A\&A, 690, A316

\bibitem[{Nelder \& Mead(1965)}]{nelder_simplex_1965}
Nelder, J. \& Mead, R. 1965, The Computer Journal, 7, 308

\bibitem[{Pacetti {et~al.}(2022)Pacetti, Turrini, Schisano, Molinari, Fonte, Politi, Hennebelle, Klessen, Testi, \& Lebreuilly}]{pacetti_chemical_2022}
Pacetti, E., Turrini, D., Schisano, E., {et~al.} 2022, ApJ, 937, 36

\bibitem[{Perrin {et~al.}(2014)Perrin, Sivaramakrishnan, Lajoie, Elliott, Pueyo, Ravindranath, \& Albert}]{perrin_updated_2014}
Perrin, M.~D., Sivaramakrishnan, A., Lajoie, C.-P., {et~al.} 2014, Proceedings of the SPIE, Volume 9143, 11 pp

\bibitem[{Phillips {et~al.}(2020)Phillips, Tremblin, Baraffe, Chabrier, Allard, Spiegelman, Goyal, Drummond, \& Hébrard}]{phillips_new_2020}
Phillips, M.~W., Tremblin, P., Baraffe, I., {et~al.} 2020, A\&A, 637, A38

\bibitem[{Rebollido {et~al.}(2024)Rebollido, Stark, Kammerer, Perrin, Lawson, Pueyo, Chen, Hines, Girard, Worthen, Ingerbretsen, Betti, Clampin, Golimowski, Hoch, Lewis, Lu, Van Der~Marel, Rickman, Seager, Soummer, Valenti, Ward-Duong, \& Mountain}]{rebollido_jwst-tst_2024}
Rebollido, I., Stark, C.~C., Kammerer, J., {et~al.} 2024, The Astronomical Journal, 167, 69

\bibitem[{Rouan {et~al.}(2000)Rouan, Riaud, Boccaletti, Clénet, \& Labeyrie}]{rouan_fourquadrant_2000}
Rouan, D., Riaud, P., Boccaletti, A., Clénet, Y., \& Labeyrie, A. 2000, PASP, 112, 1479

\bibitem[{Sanghi {et~al.}(2023)Sanghi, Liu, Best, Dupuy, Siverd, Zhang, Hurt, Magnier, Aller, \& Deacon}]{sanghi_hawaii_2023}
Sanghi, A., Liu, M.~C., Best, W. M.~J., {et~al.} 2023, The Astrophysical Journal, 959, 63

\bibitem[{Schneider {et~al.}(2023)Schneider, Munn, Vrba, Bruursema, Dahm, Williams, Liu, \& Dorland}]{schneider_astrometry_2023}
Schneider, A.~C., Munn, J.~A., Vrba, F.~J., {et~al.} 2023, The Astronomical Journal, 166, 103

\bibitem[{Shibata {et~al.}(2020)Shibata, Helled, \& Ikoma}]{shibata_origin_2020}
Shibata, S., Helled, R., \& Ikoma, M. 2020, A\&A, 633, A33

\bibitem[{Skemer {et~al.}(2016)Skemer, Morley, Zimmerman, Skrutskie, Leisenring, Buenzli, Bonnefoy, Bailey, Hinz, Defrére, Esposito, Apai, Biller, Brandner, Close, Crepp, De~Rosa, Desidera, Eisner, Fortney, Freedman, Henning, Hofmann, Kopytova, Lupu, Maire, Males, Marley, Morzinski, Oza, Patience, Rajan, Rieke, Schertl, Schlieder, Stone, Su, Vaz, Visscher, Ward-Duong, Weigelt, \& Woodward}]{skemer_leech_2016}
Skemer, A.~J., Morley, C.~V., Zimmerman, N.~T., {et~al.} 2016, ApJ, 817, 166

\bibitem[{Stolker(2023)}]{stolker_species_2023}
Stolker, T. 2023, Astrophysics Source Code Library, ascl:2307.057, aDS Bibcode: 2023ascl.soft07057S

\bibitem[{Su\'arez \& Metchev(2022)}]{suarez_ultracool_2022}
Su\'arez, G. \& Metchev, S. 2022, MNRAS, stac1205

\bibitem[{Taylor \& Atreya(2004)}]{taylor_composition_2004}
Taylor, F.~W. \& Atreya, S.~K. 2004, Jupiter. The Planet, Satellites and Magnetosphere

\bibitem[{Turrini {et~al.}(2021)Turrini, Schisano, Fonte, Molinari, Politi, Fedele, Panić, Kama, Changeat, \& Tinetti}]{turrini_tracing_2021}
Turrini, D., Schisano, E., Fonte, S., {et~al.} 2021, ApJ, 909, 40

\bibitem[{Wang {et~al.}(2021)Wang, Kulikauskas, \& Blunt}]{wang_whereistheplanet_2021}
Wang, J., Kulikauskas, M., \& Blunt, S. 2021, Astrophysics Source Code Library, record ascl:2101.003

\bibitem[{Wenger {et~al.}(2000)Wenger, Ochsenbein, Egret, Dubois, Bonnarel, Borde, Genova, Jasniewicz, Laloë, Lesteven, \& Monier}]{wenger_simbad_2000}
Wenger, M., Ochsenbein, F., Egret, D., {et~al.} 2000, A\&A Supplement Series, 143, 9

\bibitem[{Whiteford {et~al.}(2023)Whiteford, Glasse, Chubb, Kitzmann, Ray, Phillips, Biller, Palmer, Rice, Waldmann, Changeat, Skaf, Wang, Edwards, \& Al-Refaie}]{whiteford_retrieval_2023}
Whiteford, N., Glasse, A., Chubb, K.~L., {et~al.} 2023, MNRAS, 525, 1375

\bibitem[{Zahnle \& Marley(2014)}]{zahnle_methane_2014}
Zahnle, K.~J. \& Marley, M.~S. 2014, The Astrophysical Journal, 797, 41

\bibitem[{Öberg {et~al.}(2011)Öberg, Murray-Clay, \& Bergin}]{oberg_effects_2011}
Öberg, K.~I., Murray-Clay, R., \& Bergin, E.~A. 2011, ApJ, 743, L16

\bibitem[{Šubjak {et~al.}(2023)Šubjak, Lodieu, Kab\'ath, Boffin, Nowak, Grundahl, Béjar, Zapatero~Osorio, \& Antoci}]{subjak_search_2023}
Šubjak, J., Lodieu, N., Kab\'ath, P., {et~al.} 2023, A\&A, 671, A10

\end{thebibliography}

\begin{appendix}
\section{Reference Library}
\label{app:ref_lib}
\begin{table*}[h!]
    \centering
    \caption{Summary of programs used to build a reference library.}
    \begin{tabular}{c|cccccc}
    \hline
    \hline
        Programs & Filters & Star Name & Spectral type & Mag (K band) & Dithers & References\\
    \hline
        COM 1037 & F1065C & BD\,+30 2990 & K0 D & 4.7 & 9 & \cite{boccaletti_jwstmiri_2022}\\
            --   & F1140C & HD\,158896 & K5 E & 4.9 & 9 & \cite{boccaletti_jwstmiri_2022}\\
            --   & F1550C & HD\,162989 & K4III & 2.8 & 9 & \cite{boccaletti_jwstmiri_2022}\\
        ERS 1386 & F1140C, F1550C & HIP\,68245 & B2IV & 4.5 & 9 & \cite{carter_jwst_2023}\\ 
            --   &  F1065C, F1140C, F1550C & HD\,140986 & K0III & 3.6 & 5 & PI: S. Hinkley\\ 
        GTO 1277 & F1065C, F1140C & HD\,310459 & K7 E & 5.5 & 9 &\cite{malin_unveiling_2024}\\ 
        GTO 1194 & F1065C, F1140C, F1550C & HD\,218261 & F6V & 5.1 & 9 & \cite{boccaletti_imaging_2024}\\ 
        GTO 1413 & F1065C, F1140C, F1550C & HD\,190360 & G7 & 4.1 & 9 & PI: L. Pueyo\\ 
        GTO 1241 & F1065C, F1140C, F1550C & TYC\,4739-392-1& M8 & 2.9 & 9,9,5 & PI: M. Ressler\\ 
            --   & F1065C, F1140C, F1550C  &HD\,222389 & K5 & 2.9 & 5 & PI: M. Ressler \\ 
            --   & F1065C, F1140C         & CD-45\,2093 & unknown & 1.8 &  5 & PI: M. Ressler \\ 
            --   & F1065C, F1140C, F1550C & HD\,49518 & K4III & 3.8 & 5 & \cite{godoy_new_2024} \\  
        GTO 1411 & F1550C & Alpha\,Pic & A8 & 2.6 & 5 & \cite{rebollido_jwst-tst_2024}\\ 
        GO\,1668 & F1140C & * 27 Com & K4III & 1.9 & 9 & PI: S. Marino\\ 
            -- & F1140C & HD 95234 & M1III & 1.5 & 9 & PI: S. Marino\\ 
            -- & F1140C & V* BV Cap & M4III & 2.1 &  9 & PI: S. Marino\\ 
        GO\,2153 & F1140C & IRAS 17555-2235 & M6 & 3.2 & 9 & PI: G. Cugno\\ 
        GO\,2243 & F1065C, F1550C & V* DI Tuc & M5III & 1.6 & 5 & \cite{matthews_temperate_2024}\\ 
        GO\,2538 & F1140C & HD 5431 & K0  & 4.4 & 9 & PI: S. Hinkley\\ 
            -- & F1140C & HD 22333 & K0III & 4.2 & 9 & PI: S. Hinkley\\ 
        GO\,3254 & F1140C & HD 172075 & K2III & 5.2 & 9 & PI: M. Benisty\\ 
        GO\,3662& F1140C & HD 203010 & K3III & 3.5 & 9 & PI: A.-M. Lagrange\\ 
    \end{tabular}
    \tablefoot{Spectral types and magnitude value in K band come from \texttt{simbad} \citep{wenger_simbad_2000}. 
    The name of the PI from the program is indicated when no publication is yet available for the corresponding dataset.}
    \label{tab:star_ref}
\end{table*}

\section{Stellar subtraction}
\label{app:other_red}
\begin{figure*}
    \centering
    \includegraphics[width=18cm]{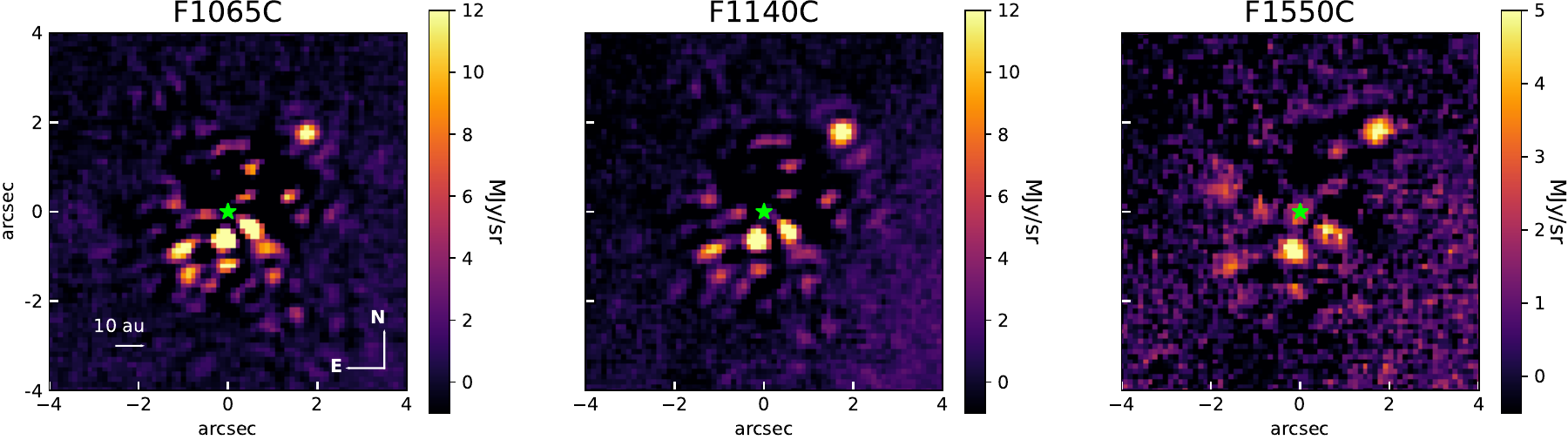}
    \includegraphics[width=18cm]{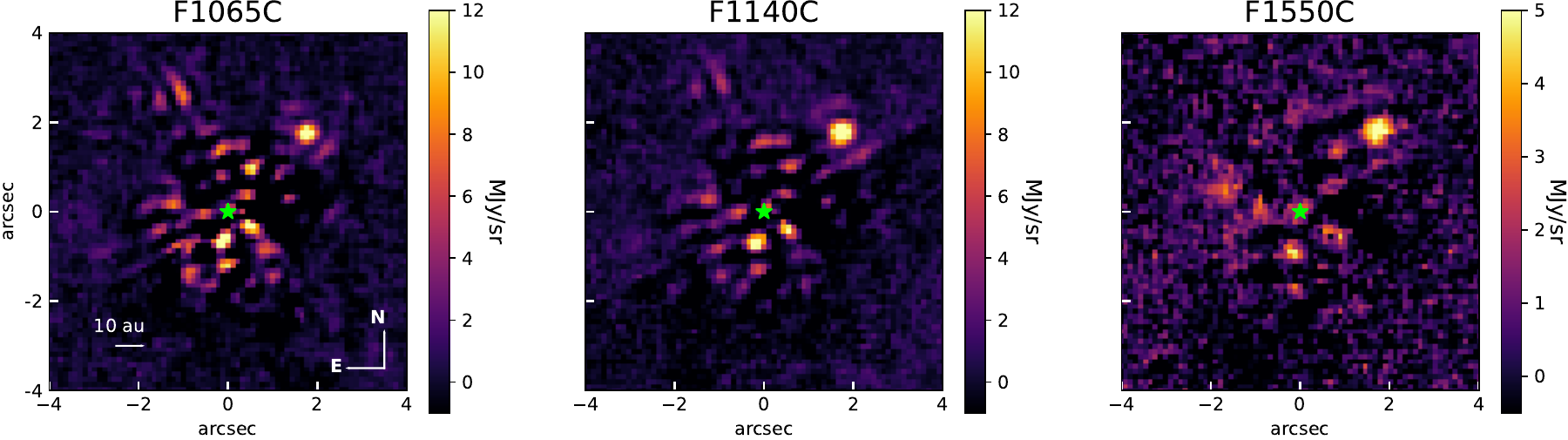}
    \caption{Subtraction of the stellar diffraction residuals, using the reference observations from the GTO 1413 only. 
    Top: Method based on PCA. 
    Bottom: Method based on PCA applied by quadrant.
    The linear optimization method gives similar results for both.}
    \label{fig:sub_gto1413}
\end{figure*}
\begin{figure*}
    \centering
    \includegraphics[width=18cm]{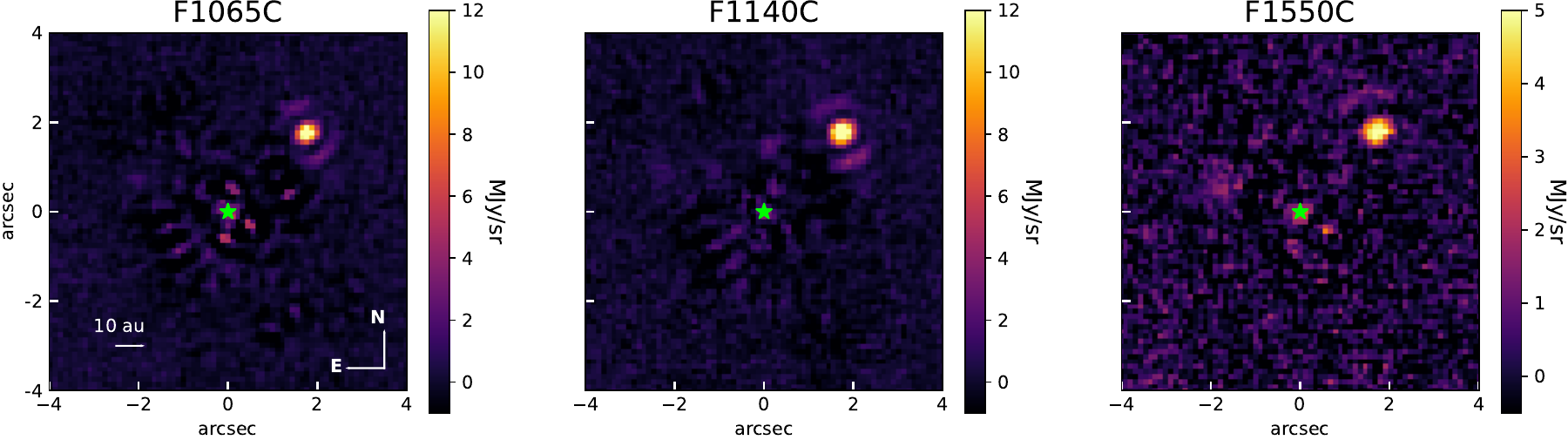}
    \includegraphics[width=18cm]{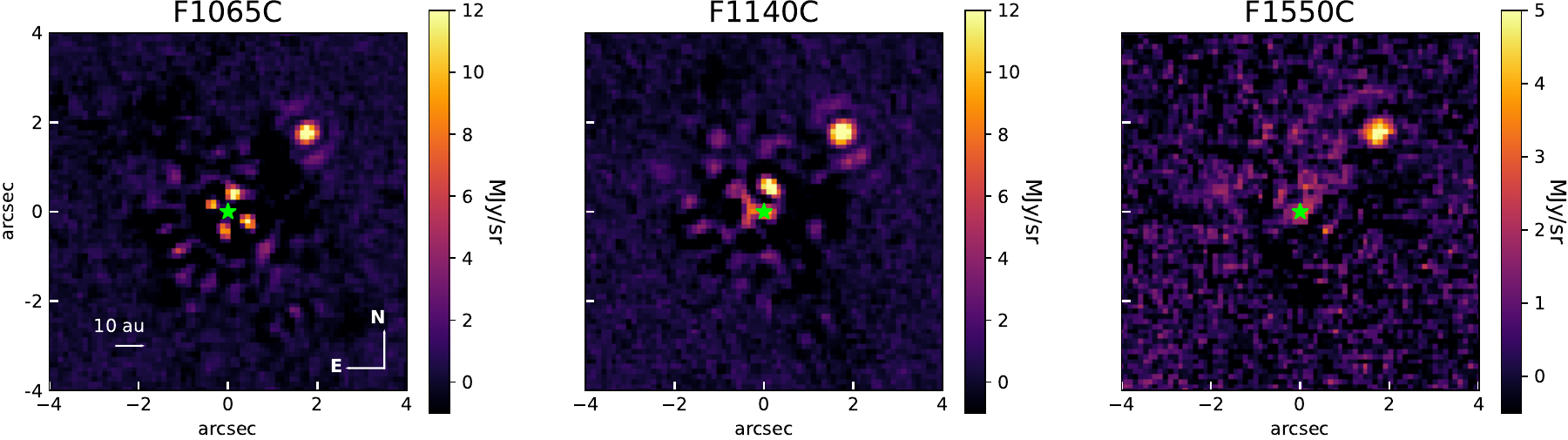}
    \caption{Subtraction of the stellar diffraction residuals, using a library of references.
    Top: Method based on classical PCA. 
    Bottom:  Method based on classical linear optimization.}
    \label{fig:sub_lib}
\end{figure*}
\begin{figure*}
    \centering
    \includegraphics[width=18cm]{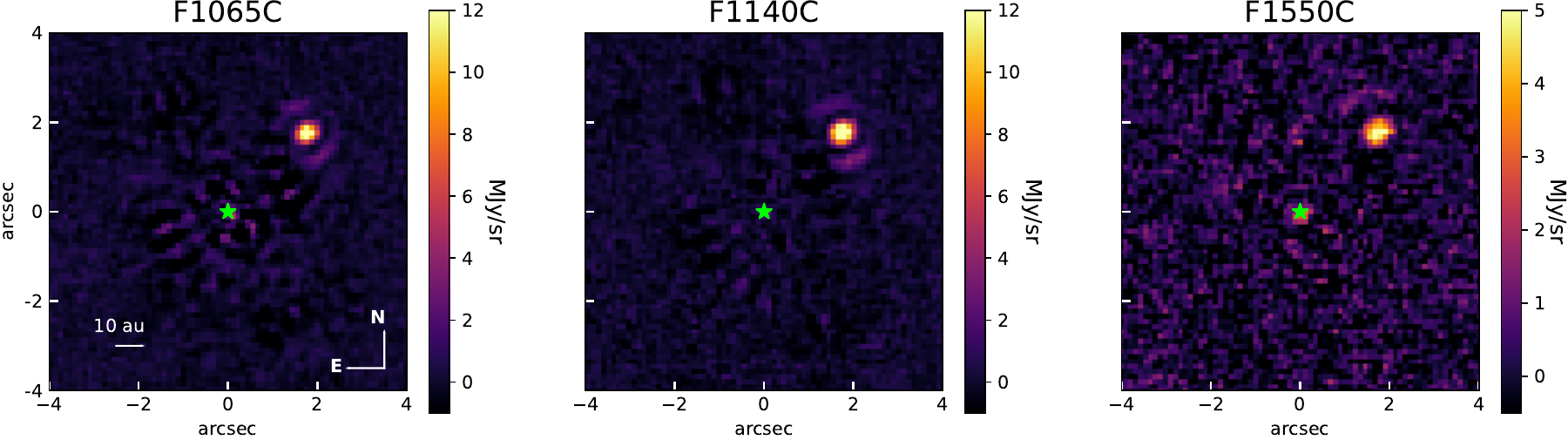}
    \includegraphics[width=18cm]{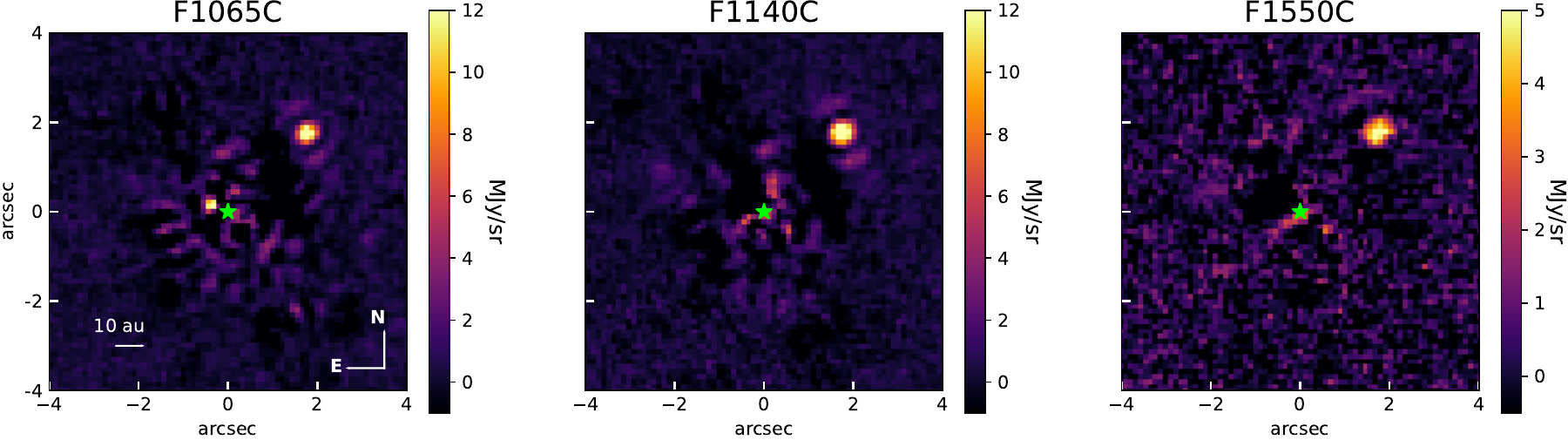}
    \caption{Subtraction of the stellar diffraction residuals, using a library of references.
    Top: Method based on PCA applied by quadrant (corresponding to the best subtraction obtained, Fig. \ref{fig:image_sub}).
    Bottom: Method based on linear optimization by quadrant.}
    \label{fig:sub_lib_4Q}
\end{figure*}

\begin{figure*}
    \centering
    \includegraphics[width=18cm]{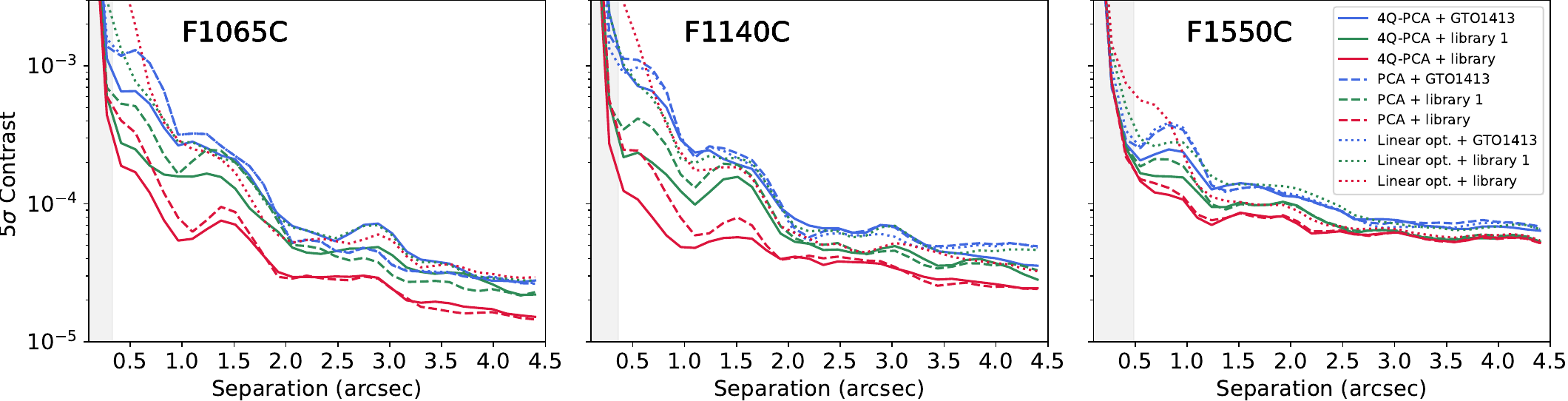}
    \caption{Contrast curves corresponding to each method, including those presented in Figs. \ref{fig:sub_gto1413}, \ref{fig:sub_lib} and \ref{fig:sub_lib_4Q}.
    The blue lines correspond to the reference star observations from the GTO 1413, the green lines from the library computed with the ERS and GTO programs, and the red lines are obtained when using the entire library (adding the GO programs available until June, 1st) — see Table \ref{tab:star_ref}.
    The lowest contrast is obtained when using the PCA method by quadrant with the library of references (dashed red line). The shaded regions correspond to the inner working angle of the coronagraphs.}
    \label{fig:contrast_curve_all}
\end{figure*}

\begin{figure}
    \centering
    \includegraphics[width=9cm]{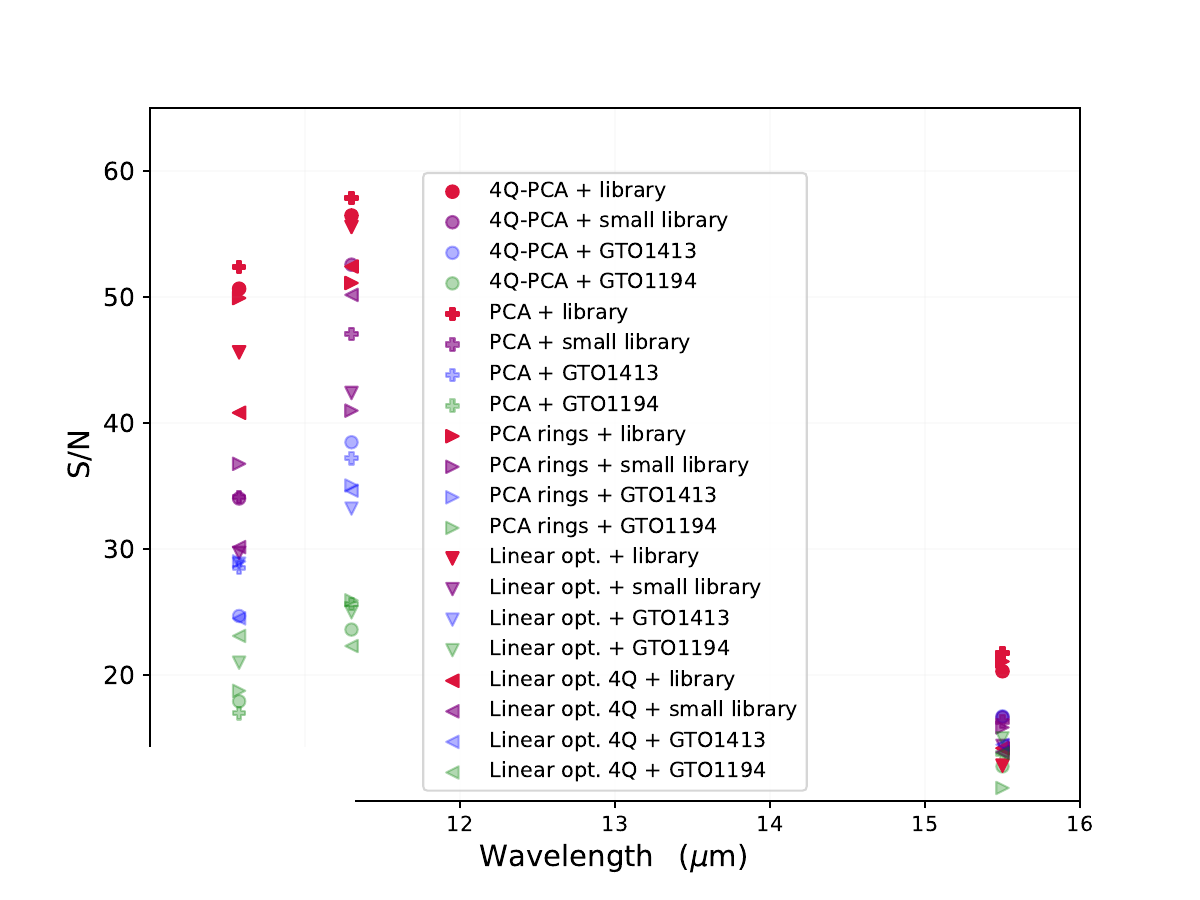}
    \caption{Values of $S/N$ for GJ\,504\,b's detection with various method for the stellar subtraction.}
    \label{fig:snr_detection}
\end{figure}

\section{Photometry measurements}
\begin{figure}
    \centering
    \includegraphics[width=9cm]{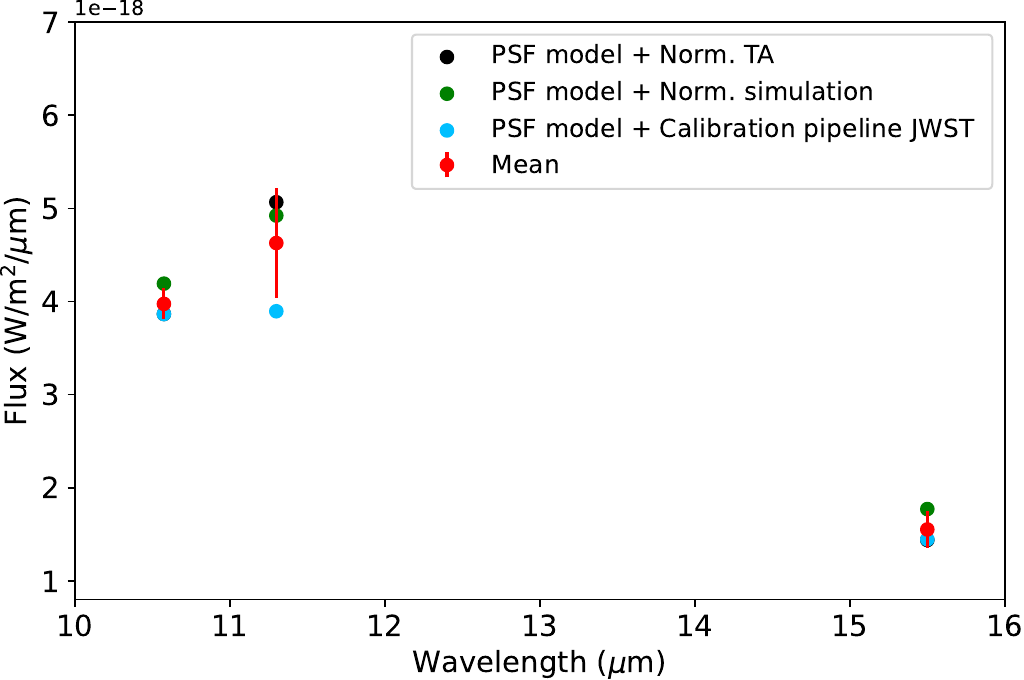}
    \caption{Photometric measurement of GJ\,504\,b using the various flux normalization methods. The red points represent the averaged of the three methods (modelling the PSF with \texttt{WebbPSF} and using TA, simulations to normalize it, or the calibration from the \texttt{JWST} pipeline).}
\label{fig:photometry_methods}
\end{figure}

\section{Model the planet's PSF}
\begin{figure*}
    \centering
    \includegraphics[width=18cm]
    {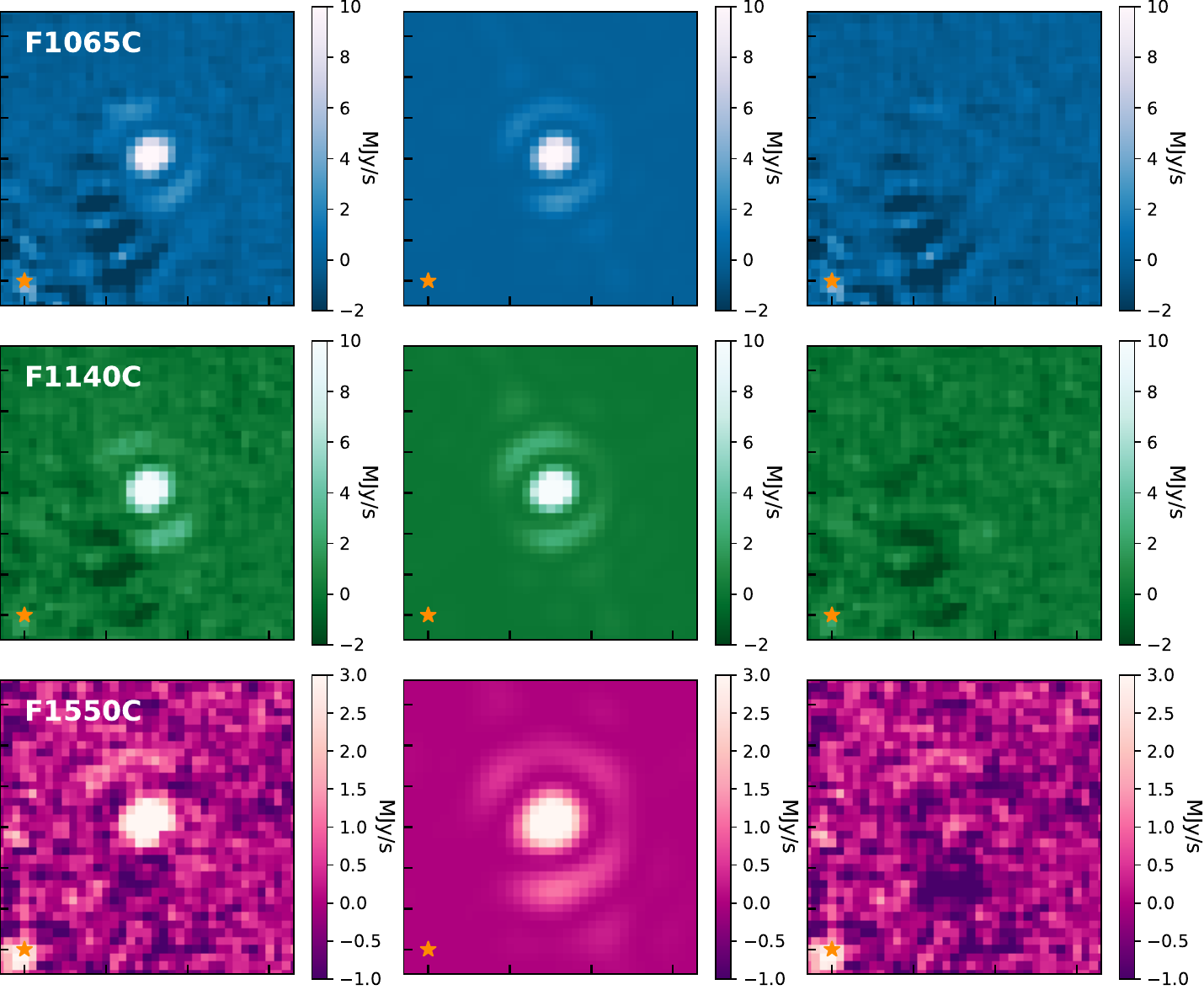}
    \caption{Left: MIRI images in each filter. Middle: \texttt{WebbPSF} model, Right: Residuals after subtracting the model from the data. The orange stars represent the center of the coronagraph masks.}
    \label{fig:model_planetl}
\end{figure*}

\section{Posterior distribution}
\begin{figure*}
    \centering
    \includegraphics[width=18cm]{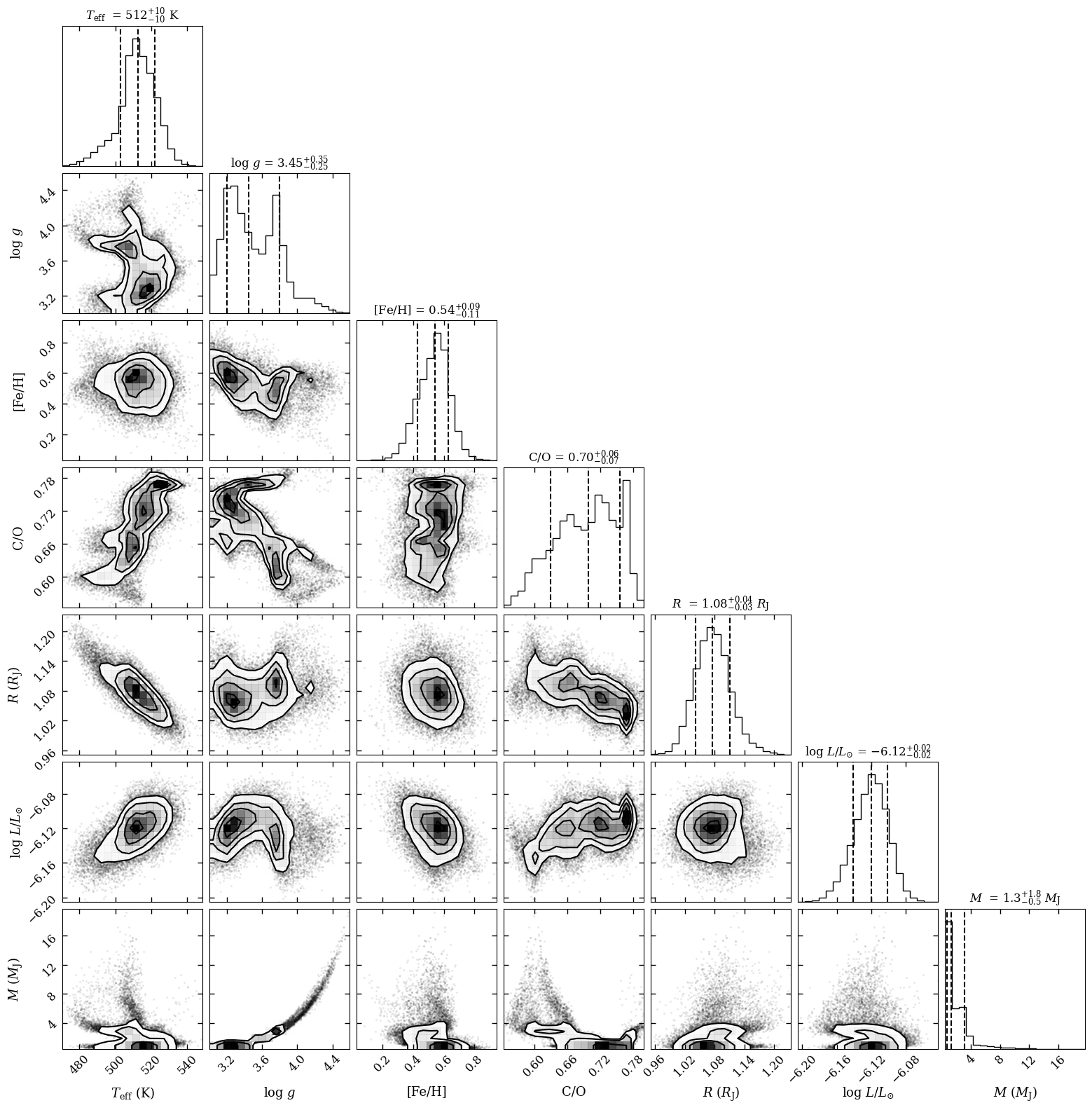}
    \caption{Posterior distribution of each atmospheric parameter from the \texttt{Exo-REM} atmospheric grids obtained when fitting the SED of GJ\,504\,b with \texttt{species}.}
    \label{fig:posterior}
\end{figure*}

\begin{figure*}
    \centering
    \includegraphics[width=18cm]{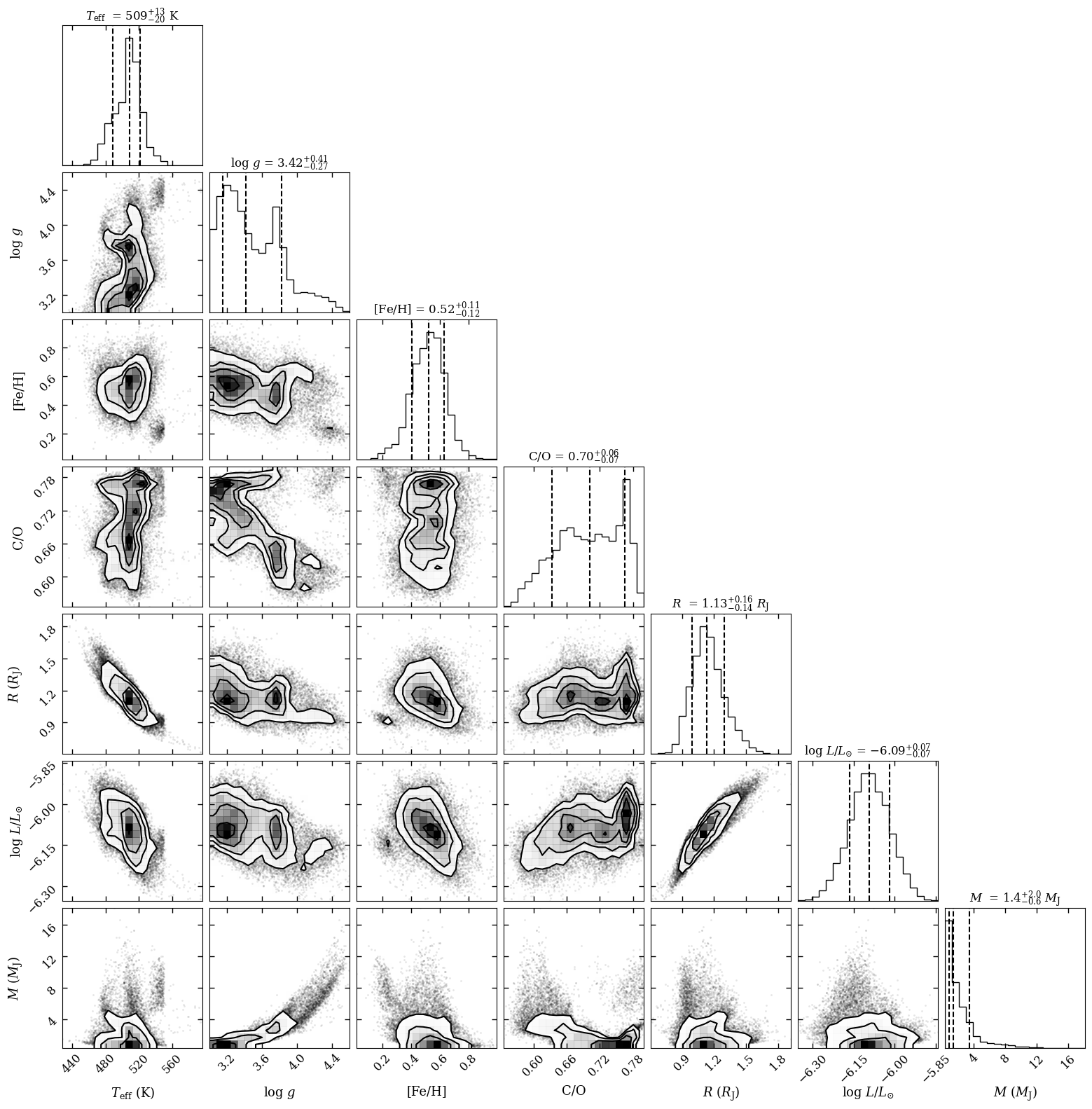}
    \caption{Posterior distributions obtained when fitting only the near-IR photometric points.}
    \label{fig:posterior_no-miri}
\end{figure*}

\section{$\Delta\chi^2$ of the NH$_3$ abundance}
\begin{figure}
    \centering
    \includegraphics[width=1\linewidth]{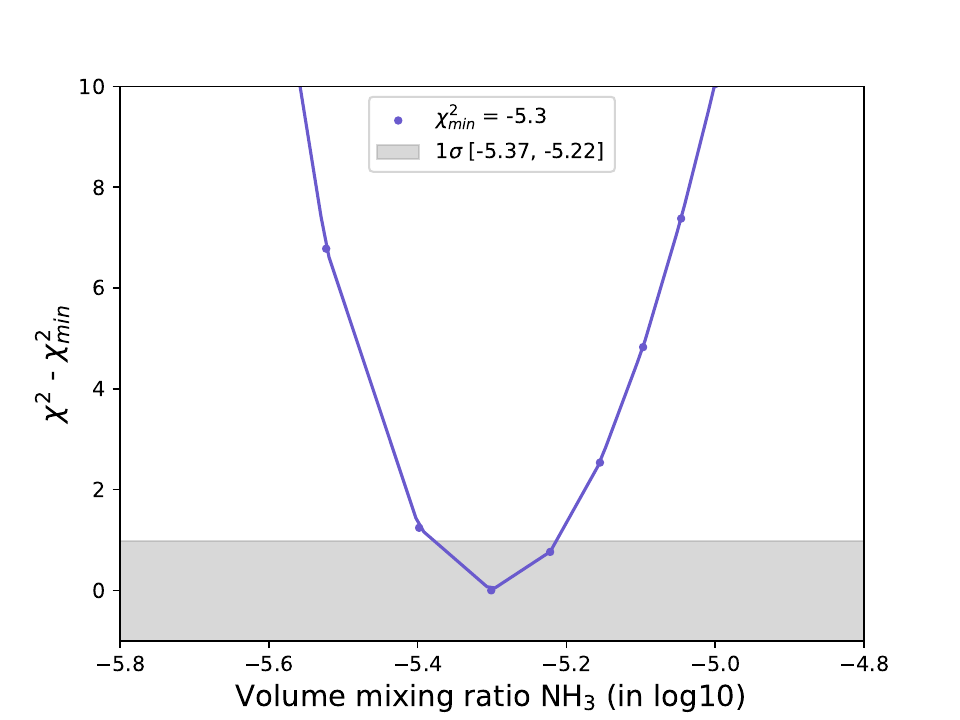}
    \caption{$\Delta\chi^2$ for each model of the \texttt{Exo-REM} atmospheric grid, varying the volume mixing ratio of NH$_3$.}
    \label{fig:chi2_min}
\end{figure}

\end{appendix}
\end{document}